\documentclass[10pt,letterpaper]{article}
\usepackage[top=0.75in,bottom=0.75in,left=0.75in,right=0.75in,footskip=0.25in]{geometry}

\usepackage[utf8]{inputenc}
\usepackage{float}
\usepackage{cite}
\usepackage{amsmath}
\usepackage{amssymb}
\usepackage{nameref,hyperref}
\hypersetup{hidelinks}
\usepackage{multirow}
\usepackage{graphicx}
\usepackage{microtype}
\DisableLigatures[f]{encoding = *, family = * }
\textwidth=\dimexpr\textwidth 
\textheight=\dimexpr\textheight
\usepackage{changepage}
\usepackage[aboveskip=1pt,font=small,labelfont=bf,labelsep=period,singlelinecheck=off]{caption}
\makeatletter
\renewcommand{\@biblabel}[1]{\quad#1.}
\makeatother
\usepackage{lastpage,fancyhdr,graphicx}
\usepackage{epstopdf}
\fancyheadoffset[L]{0in}
\fancyfootoffset[L]{0in}
\usepackage{color}
\definecolor{Gray}{gray}{.25}
\usepackage{graphicx}
\usepackage{sidecap}
\usepackage{wrapfig}
\usepackage[pscoord]{eso-pic}
\usepackage[fulladjust]{marginnote}
\reversemarginpar

\newcommand{\suppnote}[1]{%
  \refstepcounter{section}%
  \section*{Supplementary Note \arabic{section}: #1}%
  \phantomsection%
  \addcontentsline{toc}{section}{Supplementary Note \arabic{section}: #1}%
}
\begin{document}
\vspace*{0.35in}

\begin{flushleft}
{\Large
\textbf\newline{Cell size and confinement drive asymmetric cell division through a cortical instability}
}
\newline
\\
Da Gao\textsuperscript{1,\textsection},
Guoye Guan\textsuperscript{2,\textsection,\textdagger,*},
Chao Tang\textsuperscript{2,3,\textdaggerdbl},
Rui Ma\textsuperscript{1,4,*}
\\
\bigskip
\bf{1} Department of Physics, College of Physical Science and Technology, Xiamen University, Xiamen 361005, People’s Republic of China
\\
\bf{2} Center for Quantitative Biology, Peking University, Beijing 100871, People’s Republic of China
\\
\bf{3} Peking-Tsinghua Center for Life Sciences, Academy for Advanced Interdisciplinary Studies, Peking University, Beijing 100871, People’s Republic of China
\\
\bf{4} Fujian Provincial Key Laboratory for Soft Functional Materials Research, Research Institute for Biomimetics and Soft Matter, Xiamen 361005, People’s Republic of China
\\
\bigskip
\textsection These authors contributed equally to this work.
\\
\textdagger Current Address: Department of Systems Biology, Harvard Medical School, Boston 02115, United States of America; Department of Data Science, Dana-Farber Cancer Institute, Boston 02215, United States of America
\\
\textdaggerdbl Current Address: Center for Interdisciplinary Studies, Westlake University, Hangzhou 310030, People’s Republic of China
\\
* Corresponding authors: ruima@xmu.edu.cn, guanguoye@gmail.com

\end{flushleft}

\textbf{Asymmetric cell division — in which a mother cell divides into two daughter cells of unequal size — is a fundamental problem in biology. It is believed that the asymmetry originates from the prior polarization of the mother cell. Here we show that division asymmetry can occur spontaneously even in unpolarized mother cells. Specifically, curvature-dependent active stresses in the cell cortex can lead to this symmetry breaking without any molecular polarity cue if the mother cell is confined within a restricted space. Either reducing the cell size or tightening mechanical confinement triggers the same spontaneous symmetry-breaking instability, in which the contractile ring slips off the equator to yield daughters of unequal volume. In the presence of a polarity cue, this instability cooperates with the cue to program the division asymmetry. The model prediction is compared with the imaging data of \textit{C.\,elegans} embryogenesis, in which successive cell divisions in a confined eggshell lead to smaller and smaller cell sizes. The measured division asymmetry indeed increases as the cells shrink, and is further amplified when the embryo is mechanically compressed, both in agreement with the model prediction.}

\vspace{1em}

Cell division is executed by the actomyosin cortex---a thin, contractile gel beneath the plasma membrane---whose equatorial constriction partitions the cytoplasm into two daughters~\cite{salbreux2012,pollard2003,chugh2017}. Whether this partition is symmetric or asymmetric in volume determines cell fate, developmental timing, and tissue architecture~\cite{neumuller2009,mukherjee2015}. In \textit{Caenorhabditis elegans}, the germline lineage divides asymmetrically while the somatic founder cells it produces divide nearly symmetrically, and the degree of germline size asymmetry follows a precise, reproducible pattern across successive divisions~\cite{eLife2021,arata2010}. What physically sets this asymmetry, and how it responds to cell size and mechanical environment, remains an open question.

Three-dimensional time-lapse imaging has produced detailed morphological atlases of these cleavages~\cite{fickentscher2016,eLife2021,POWERS2010}, which reveal a stereotyped pattern of division asymmetry that most accounts attribute to the anterior--posterior polarity established by PAR (partitioning defective) proteins~\cite{kemphues1988,goehring2011,sailer2015}: polarity biases cortical contractility and thereby shifts the cleavage plane away from the position set by pure geometry~\cite{eLife2021}. Embryonic blastomeres, however, rarely divide in isolation. Cleavage proceeds inside a rigid eggshell, and each blastomere is further hemmed in by its neighbours as the embryo subdivides, so that every division takes place within a mechanically confined space~\cite{fickentscher2013,tian2020}. The blastomeres themselves also become progressively smaller with each successive round. The contribution of these geometric factors---cell size and external mechanical confinement---to the degree of size asymmetry has received comparatively little attention.

Active surface theory provides a natural framework for describing the dividing cortex~\cite{salbreux2012,turlier2014,julicher2018,salbreux2017}. Within this theory, surfaces fall into distinct classes according to which mirror symmetries they break, and the class dictates which active stresses are admissible~\cite{salbreux2017}. Most cortex models belong to the simplest class, which preserves all mirror symmetries and permits only an isotropic active tension. Such surfaces polarize spontaneously rather than furrow~\cite{mietke2019self,Gao2026}, so that generating a cleavage furrow within this class requires the contractile ring to be imposed by hand, through a prescribed enrichment of myosin at the equator~\cite{turlier2014,Gao2026}. The cortex, however, belongs to a different class, in which the up-down mirror symmetry is broken. Such breaking arises naturally, either from the asymmetric environment of the cortex, whose two faces are bounded by the plasma membrane on one side and by the cytoplasm on the other, or from a gradient of myosin activity across the finite cortical thickness~\cite{berthoumieux2014} (Fig.~\ref{fig:mechanical_model}A, inset; see Discussion). Either source admits active stresses that are linear in the curvature tensor~\cite{salbreux2017}, and the shape dynamics they generate have remained largely unexplored.

Here we model the cortex as a closed viscous surface enclosed by an ellipsoidal shell, retaining the curvature-dependent active stresses that its broken up-down symmetry admits and allowing the bending rigidity to be modulated by a polarity cue. No contractile ring is imposed by hand. The cleavage furrow instead emerges from the curvature-dependent active stresses acting on an ellipsoidal cortex whose curvature gradients are set by the confinement. Using dynamic simulations together with numerical continuation of the steady-state branches, we map the division outcomes across cell size and degree of confinement, and test the resulting predictions against three-dimensional time-lapse imaging of wild-type \textit{C.\,elegans} embryos.

\section*{Curvature-dependent active surface model}

\begin{figure}[htbp]
\includegraphics[width=\textwidth]{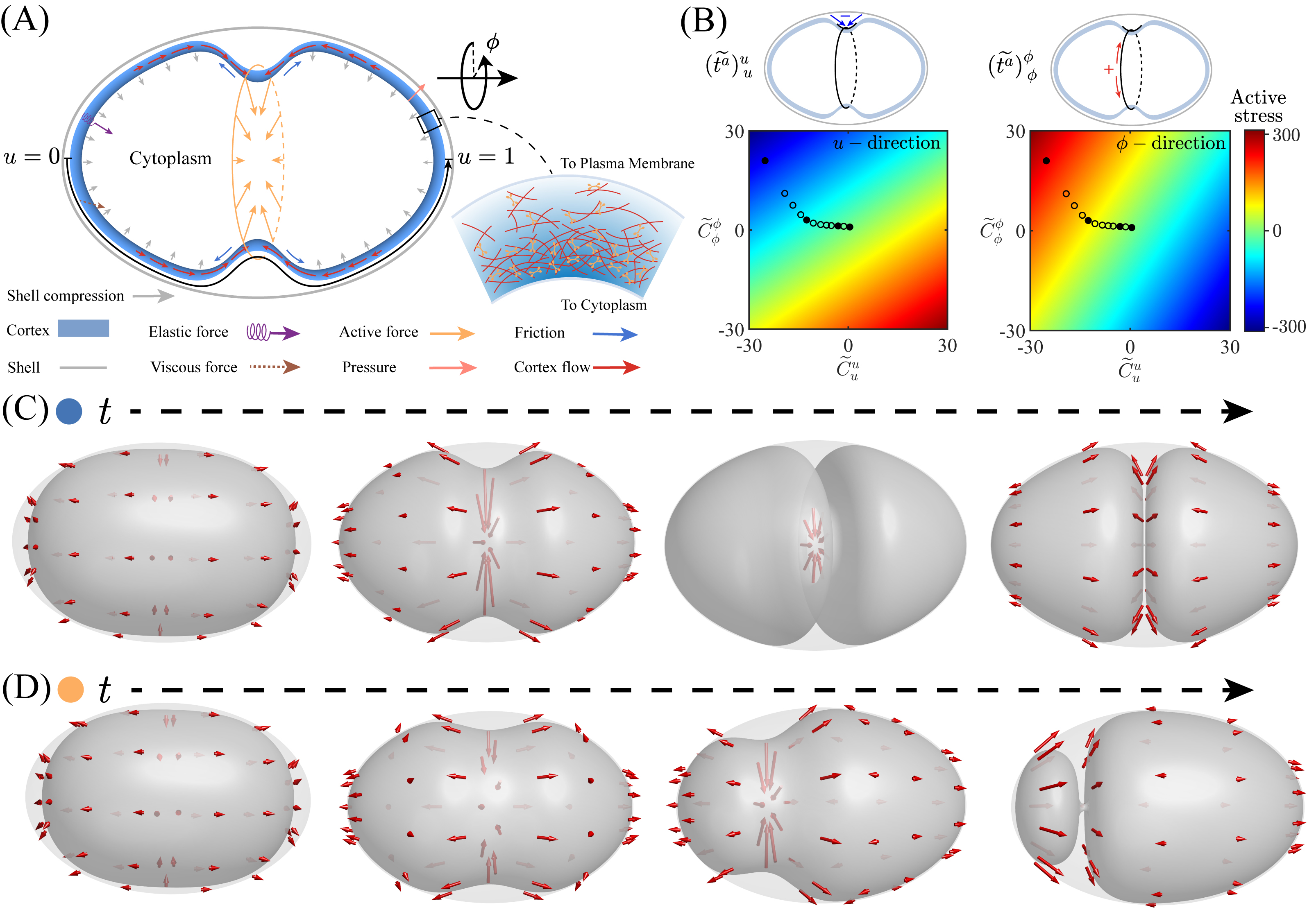}
\caption{\textbf{Curvature-guided active surface model for cell division under spatial confinement.} \textbf{A}, Axisymmetric active-surface model: the cortex is a closed viscoelastic surface $\mathbf{X}(u,\phi,t)$ inside an ellipsoidal confining shell (semi-transparent gray), with shape and flow $\mathbf{v}$ set by the force balance of Eq.~(\ref{eq:forcebalance}). The color gradient across the cortex thickness marks its broken up-down symmetry (cytoplasm-facing, dark; membrane-facing, light). Inset: cross-section showing the actin network (red) and myosin motors (orange), whose density increases toward the cytoplasmic face, a possible microscopic origin of the curvature-dependent active stress (see Discussion). \textbf{B}, Active stress components $(t^a)^u_{\,u}$, $(t^a)^\phi_{\,\phi}$ in the principal-curvature plane $(C^u_u,C^\phi_\phi)$. Dots trace the furrow curvature during the symmetric division of \textbf{C}, the four black solid dots marking the four snapshots shown there. \textbf{C,D}, Snapshots of a symmetric (\textbf{C}) and asymmetric (\textbf{D}) division. Red arrows, active force density $\mathbf{f}^a=\mathrm{div}(\mathbf{T}^a)$. In the third snapshot of \textbf{C}, arrows appear only near the equator because $\mathbf{f}^a$ is elsewhere too small to display.}
\label{fig:mechanical_model}
\end{figure}

Because the actin filaments and myosin motors of the cortex turn over on timescales of seconds---far shorter than the minutes-long division process---the cortex behaves as a viscous fluid on the timescale of cytokinesis while resisting bending as an elastic film~\cite{turlier2014}. We therefore model it as a closed, axisymmetric viscous surface enclosed by an ellipsoidal shell that represents the mechanical confinement of the embryonic environment (Fig.~\ref{fig:mechanical_model}A). Cortical flows follow from force balance, $\mathrm{div}(\mathbf{T})=-\mathbf{f}^{\mathrm{ext}}$, with the total stress $\mathbf{T}$ comprising Helfrich bending elasticity, viscous dissipation, and the active stress generated by myosin motors (Methods).

The active stress is the ingredient specific to this work. Because the up-down symmetry of the cortex is broken, the theory admits active stresses that are linear in the curvature tensor~\cite{salbreux2017},
\begin{equation}
t^a_{ij}=\xi_s\Delta\mu\!\left(C_{ij}-\tfrac{1}{2}C^k_k g_{ij}\right)+\xi_b\Delta\mu\,C^k_k g_{ij}\label{eq:act},
\end{equation}
where $\Delta\mu$ is the chemical energy released per ATP hydrolysis cycle, and $\xi_s\Delta\mu$ and $\xi_b\Delta\mu$ quantify how efficiently the motors couple this output to the deviatoric and isotropic components of the cortical curvature. For an axisymmetric surface, the $\xi_s$ term is proportional to the principal curvature difference $C^\phi_\phi - C^u_u$ and the $\xi_b$ term to the mean curvature $C^k_k = C^\phi_\phi + C^u_u$. A cell confined within an ellipsoidal shell adopts an ellipsoidal resting shape in which $C^\phi_\phi - C^u_u$ is maximal at the equator and $C^k_k$ is larger near the poles. The $\xi_s$ term therefore concentrates anisotropic contractile stress at the equator, while the $\xi_b$ term generates extensile isotropic stress near the poles (Fig.~\ref{fig:mechanical_model}B,C). Throughout, cell size is measured by the reduced mother-cell radius $\tilde{R}_0=R_0/L_h$, with $L_h$ the hydrodynamic length, and confinement by the volume ratio $\epsilon=V_\mathrm{cell}/V_\mathrm{shell}$ (Methods).

Figure~\ref{fig:mechanical_model}C,D presents the two characteristic dynamics of cortical shape evolution predicted by the model. In the symmetric case (Fig.~\ref{fig:mechanical_model}C; Supplementary Movie~1), the cortex changes from a smooth ellipsoid through a shallow equatorial groove into two nearly spherical lobes joined by an increasingly narrow neck, the furrow tightening symmetrically about the equator. In the asymmetric case (Fig.~\ref{fig:mechanical_model}D; Supplementary Movie~2), mirror symmetry is spontaneously broken and the furrow forms off the equator, migrating toward one pole to yield daughter cells of unequal volume. The mechanical origin of the symmetric furrowing lies in the principal curvatures at the furrow. Mapping the two active stress components $(t^a)^u_{\,u}$ and $(t^a)^\phi_{\,\phi}$ in the principal-curvature plane $(C^u_u,C^\phi_\phi)$ (Fig.~\ref{fig:mechanical_model}B) and tracking the furrow curvature through successive stages (dots) shows that as the furrow deepens $C^u_u$ becomes increasingly negative and $C^\phi_\phi$ increasingly positive, making $(t^a)^\phi_{\,\phi}$ progressively more contractile and driving the ring to narrow against bending resistance. The resulting active force density $\mathbf{f}^a=\mathrm{div}(\mathbf{T}^a)$ (red arrows, Fig.~\ref{fig:mechanical_model}C,D) concentrates ever more strongly at the constriction as it narrows, enabling an extremely deep cleavage furrow. This positive curvature–stress feedback sustains both trajectories, whose normal force balance, decomposed into its active, viscous and elastic contributions, is detailed in Supplementary Fig.~S2.

These two outcomes—symmetric and asymmetric division—represent qualitatively distinct solutions of the same active-surface equations. We next explore systematically how the two geometric parameters, $\tilde{R}_0$ and $\epsilon$, select between them (Fig.~\ref{fig:size_confinement}), showing that both can trigger the underlying symmetry-breaking instability.

\section*{Size and confinement control division symmetry}
\begin{figure}[htbp]
\includegraphics[width=\textwidth]{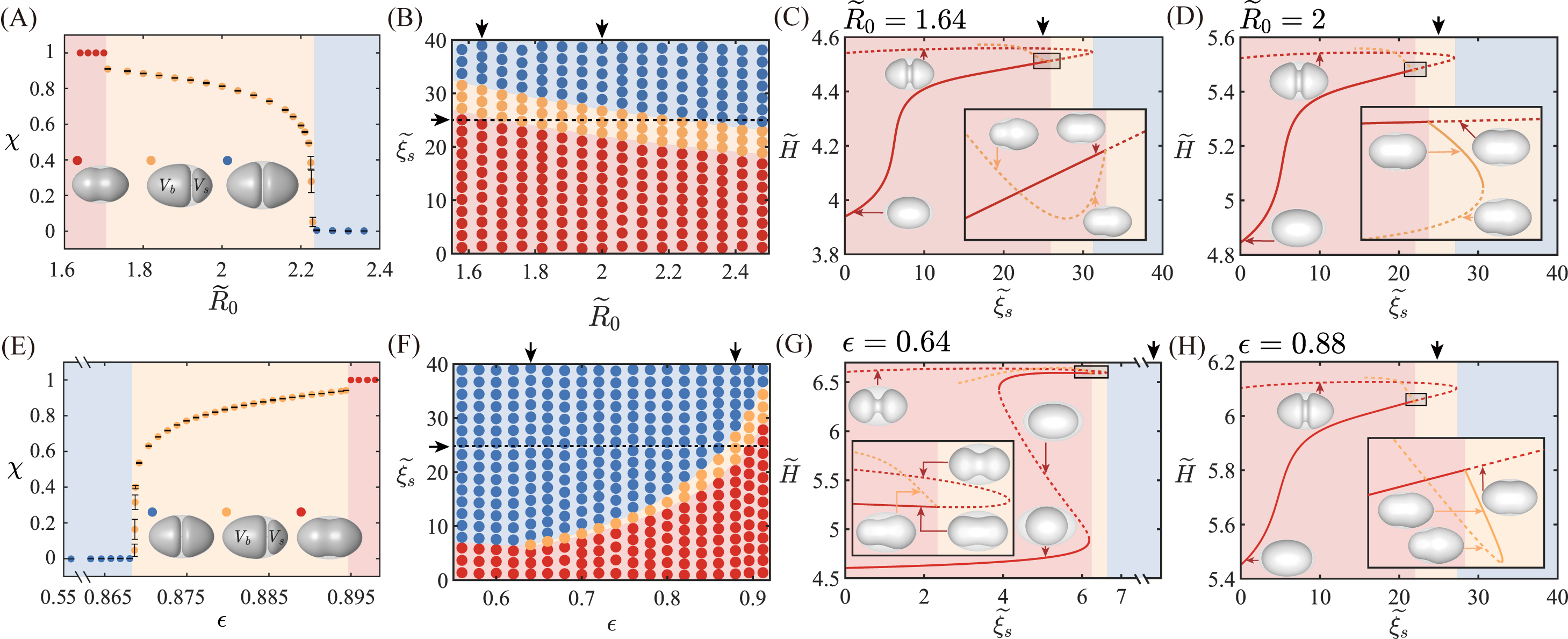}
\caption{\textbf{Cell size and confinement both trigger a symmetry-breaking instability.} Panels \textbf{A}--\textbf{D} vary cell size at fixed confinement ($\epsilon=0.87$), panels \textbf{E}--\textbf{H} vary confinement at fixed size ($\tilde{R}_0=2.22$). \textbf{A,E}, Division asymmetry $\chi$ versus cell radius $\tilde{R}_0$ (\textbf{A}) and shell confinement $\epsilon$ (\textbf{E}), with the three regimes colored symmetric ($\chi\approx 0$, blue), asymmetric (orange) and failed division (red). Failed divisions are assigned $\chi=1$, as no partition is completed. Error bars, standard deviation over ten simulations, each with a different random perturbation of the active stress applied at the first time step (Supplementary Information). \textbf{B,F}, Phase diagrams in the $(\tilde{R}_0,\tilde{\xi}_s)$ and $(\epsilon,\tilde{\xi}_s)$ planes. The two arrows above mark the parameter values examined in the panels to the right, and the arrow on the left, with the dashed horizontal line, marks the activity $\tilde{\xi}_s$ used in \textbf{A} and \textbf{E}. \textbf{C,D,G,H}, Steady-state bifurcation landscapes ($\tilde{H}$ vs.\ $\tilde{\xi}_s$) at $\tilde{R}_0=1.64$ and $2.00$ (\textbf{C,D}) and $\epsilon=0.64$ and $0.88$ (\textbf{G,H}). Solid and dashed branches denote stable and unstable solutions, insets show representative cortex shapes, and the arrow above each panel marks the same activity as in \textbf{A,E}. For each of these four cases, the near-threshold scaling of the contractile-ring displacement is given in Supplementary Fig.~S6.}
\label{fig:size_confinement}
\end{figure}

To clarify how cell size governs the symmetry of division, we perform dynamic simulations for various cell sizes $\tilde{R}_0$ at fixed confinement $\epsilon=0.87$, so that the shell volume is reduced in proportion to the cell as $\tilde{R}_0$ decreases. Three distinct outcomes emerge (Fig.~\ref{fig:size_confinement}A). For large cells, a constriction furrow nucleates at the equatorial plane and narrows symmetrically, yielding daughter cells of equal volume, quantified by the division asymmetry index $\chi=(V_b-V_s)/(V_b+V_s)\approx 0$, where $V_b$ and $V_s$ are the volumes of the bigger and smaller daughter cells. At intermediate $\tilde{R}_0$, the furrow likewise initiates at the equator but, as it narrows, its position---the locus where the meridional curvature $C^u_u$ is most negative---spontaneously slips toward one pole, yielding daughters of unequal volume with $\chi$ increasing continuously as $\tilde{R}_0$ decreases. For very small cells, the cortex forms only a weakly invaginated furrow and fails to constrict further, arresting cytokinesis (Supplementary Movie~3). The phase diagram of Fig.~\ref{fig:size_confinement}B shows that decreasing $\tilde{R}_0$ progressively raises the symmetry-breaking threshold $\tilde{\xi}_s^*$ and expands the failed-division region, so that smaller cells require stronger motor activity to break equatorial symmetry but, once the threshold is exceeded, produce greater division asymmetry.

To resolve the underlying mechanism, we use numerical continuation to compute the steady-state solution branches as a function of $\tilde{\xi}_s$ at two representative radii, $\tilde{R}_0=1.64$ and $2.00$ (Fig.~\ref{fig:size_confinement}C,D; Methods). Two classes of solutions are distinguished by their symmetry: a red branch of geometries that preserve mirror symmetry about the equatorial plane, which at $\tilde{\xi}_s=0$ adopts the ellipsoidal shape imposed by the confining shell, and an orange branch of geometries whose furrow is displaced off the equator. At a critical activity $\tilde{\xi}_s^*$ the symmetric branch loses stability and the asymmetric branch bifurcates from it. At still higher activity the symmetric branch reaches a turning point and folds back, beyond which no symmetric steady state exists. These two thresholds partition the $\tilde{\xi}_s$ axis into exactly the three regimes found dynamically in Fig.~\ref{fig:size_confinement}A. Below the bifurcation the symmetric branch is stable and the cortex settles onto it, giving failed division. Between the bifurcation and turning points the symmetric branch is unstable, so mirror symmetry breaks and division is asymmetric. Beyond the turning point no symmetric steady state remains and the cortex constricts symmetrically until the furrow radius becomes extremely small. At the activity used in Fig.~\ref{fig:size_confinement}A, $\tilde{R}_0=1.64$ lies below the bifurcation and $\tilde{R}_0=2.00$ within the asymmetric window, reproducing the failed and asymmetric outcomes respectively.

Near threshold, the displacement of the ring from the equator, $\Delta u = u_\mathrm{furrow}-0.5$, follows the square-root scaling $|\Delta u|\propto|\tilde{\xi}_s-\tilde{\xi}_s^*|^{1/2}$ characteristic of a pitchfork bifurcation (Supplementary Information, Figs.~S3 and~S6A,B). Its character changes with cell size: the transition is supercritical at $\tilde{R}_0=2.00$ and subcritical at $\tilde{R}_0=1.64$.

The same analysis along the confinement axis yields the same three-regime structure. Computing $\chi$ against $\epsilon$ at fixed $\tilde{R}_0=2.22$ gives symmetric division at weak confinement, a continuous rise in $\chi$ above a critical $\epsilon$, and division failure beyond $\epsilon_c\approx 0.895$ (Fig.~\ref{fig:size_confinement}E), and the phase diagram of Fig.~\ref{fig:size_confinement}F shows that tightening confinement raises $\tilde{\xi}_s^*$ and expands the failed-division region exactly as decreasing $\tilde{R}_0$ does. The branch structure and its correspondence with the dynamics are likewise unchanged (Fig.~\ref{fig:size_confinement}G,H). Two features are specific to this axis. First, the bifurcation character shifts in the opposite sense, being supercritical at $\epsilon=0.88$ and subcritical at $\epsilon=0.64$ (Supplementary Fig.~S6C,D), whereas along the size axis it becomes subcritical as the cell shrinks. Second, at weak confinement the symmetric branch acquires an S-shaped structure, so that an ellipsoidal state without ingression and an already-furrowed state coexist as stable solutions over a range of $\tilde{\xi}_s$ (Fig.~\ref{fig:size_confinement}G). At an even weaker confinement, $\epsilon=0.56$, the two folds of this S swap their order along $\tilde{\xi}_s$, so that increasing activity carries the cortex from failed division directly to symmetric constriction, eliminating the asymmetric-division window altogether (Supplementary Fig.~S4).

Cell size and confinement therefore act on one and the same active-surface instability, providing a unified mechanical basis for how geometric cues set the degree of division asymmetry.

\section*{Polarity regulation biases the division plane}
\begin{figure}[htbp]
\includegraphics[width=\textwidth]{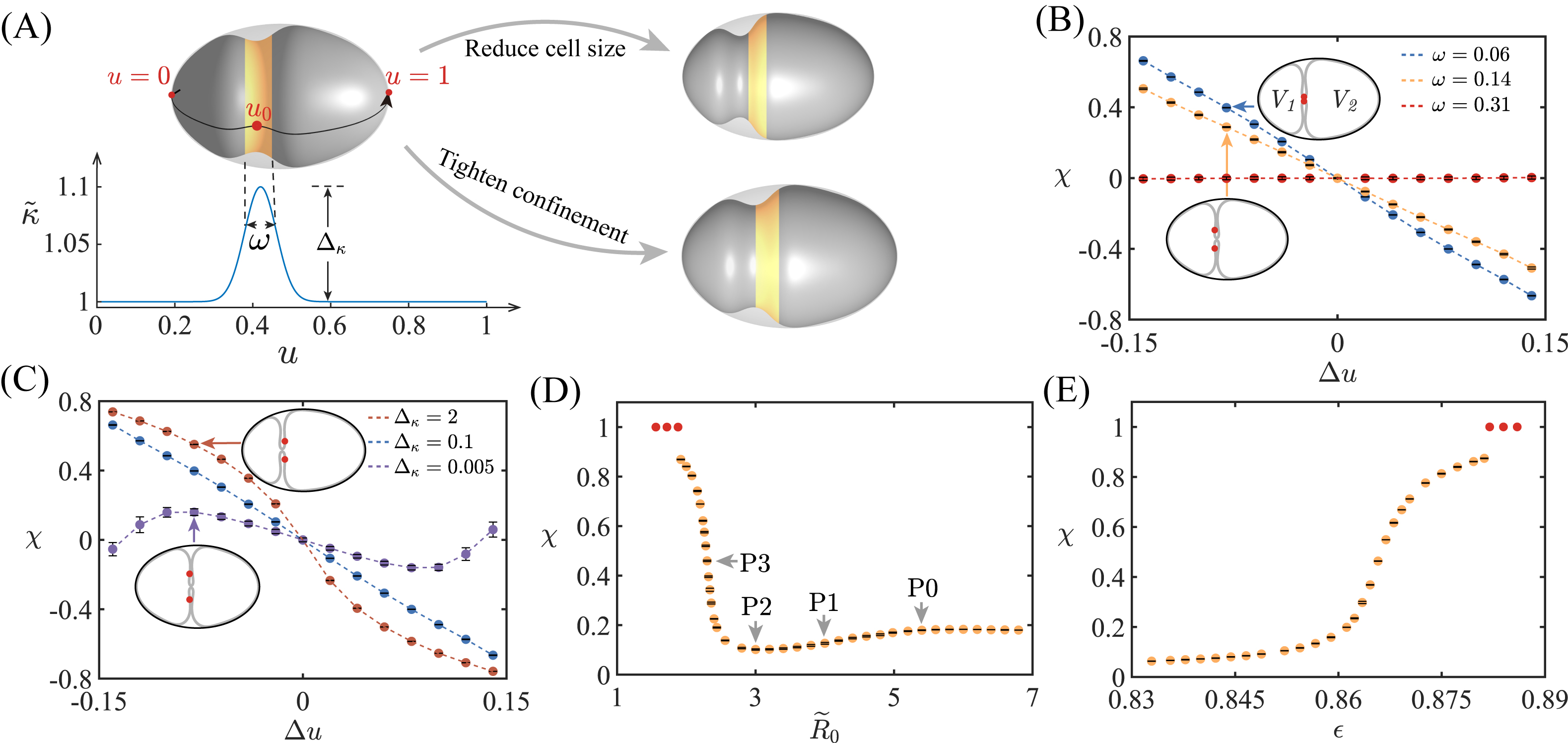}
\caption{\textbf{Localized polarity regulation specifies the position and magnitude of asymmetric division.} \textbf{A}, Left: bending-rigidity profile $\tilde{\kappa}(u)=1+\Delta_\kappa\exp[-(u-u_0)^2/\omega^2]$, a Gaussian pulse of width $\omega$ and amplitude $\Delta_\kappa$ centered off the equator at $u_0$. Right: cortex shapes for reduced cell size (upper) and reduced shell size (lower). \textbf{B,C}, Division asymmetry $\chi$ versus the polarity offset $\Delta u = u_0 - 0.5$ at $\tilde{R}_0=5.4$, for profile widths $\omega=0.06,\,0.14,\,0.31$ (\textbf{B}) and amplitudes $\Delta_\kappa=0.005,\,0.1,\,2$ (\textbf{C}). Insets: cortex shapes at the marked points, where both red dots mark the prescribed cue position $u_0$. In \textbf{B} the furrow coincides with $u_0$ for $\omega=0.06$ (upper) but is displaced from it for $\omega=0.14$ (lower). In \textbf{C} a strong cue ($\Delta_\kappa=2$, upper) drives the furrow beyond $u_0$, producing an asymmetry stronger than $u_0$ alone would set, whereas a weak cue ($\Delta_\kappa=0.005$, lower) leaves the furrow short of $u_0$, producing a weaker one. \textbf{D,E}, $\chi$ versus cell radius $\tilde{R}_0$ at fixed $\epsilon=0.87$ (\textbf{D}, arrows mark the P0--P3 radii), and versus shell confinement $\epsilon$ at fixed $\tilde{R}_0=2.22$ (\textbf{E}), both with $u_0=0.465$, $\omega=0.06$ and $\Delta_\kappa=0.1$. Red dots, division failure. Error bars in \textbf{B}--\textbf{E} as in Fig.~\ref{fig:size_confinement}.}
\label{fig:regulation}
\end{figure}

Embryonic cytokinesis is typically also guided by an intrinsic polarity cue. To incorporate one while preserving the mechanical responses identified above, we modulate the cortical bending rigidity locally,
\begin{equation}
\kappa(u)=\kappa_0\left[1+\Delta_{\kappa}\exp\!\left(-\frac{(u-u_0)^2}{\omega^2}\right)\right]\!,
\end{equation}
a Gaussian pulse centered at $u_0$ with width $\omega$ and amplitude $\Delta_\kappa$ (Fig.~\ref{fig:regulation}A). Modulating $\kappa$ is mechanistically motivated: the contractile ring is enriched in actin filaments and crosslinking proteins relative to the surrounding cortex~\cite{PRL2009,salbreux2012}, so the bending rigidity is locally elevated at the ring, and both crosslinker density and actin network architecture are downstream targets of PAR polarity proteins in \textit{C.\,elegans}~\cite{James2004,reymann2016,jcb2016}. How faithfully the cue encodes a division plane depends on its width and amplitude. For narrow profiles the daughter-cell asymmetry $\chi$ varies near-linearly with the cue offset $\Delta u$, so that small displacements translate into proportional changes of $\chi$, whereas broader profiles flatten the relation and decouple cue position from division outcome (Fig.~\ref{fig:regulation}B). The amplitude controls the sensitivity of the response (Fig.~\ref{fig:regulation}C): below a threshold the cortex fails to track the cue, and at large $\Delta_\kappa$ the response saturates. Narrow profiles of intermediate amplitude therefore define the operating window for high-fidelity polarity encoding.

Coupling the polarity profile to the geometric control parameters produces two distinct trends. Along the cell-size axis, $\chi$ approaches a constant set by the cue offset $u_0$ alone for large cells, and rises sharply once the cell-size induced instability of Fig.~\ref{fig:size_confinement}A takes over at small $\tilde{R}_0$ and progressively amplifies the polarity-induced asymmetry (Fig.~\ref{fig:regulation}D). The arrows mark the radii of the four successive germline blastomeres, set recursively from the P0 zygote, whose size is the single calibrated input: each germline division produces two daughters and the smaller one becomes the next germline cell, so that the arrows track the shrinking lineage along the curve. We show below that this predicted escalation is qualitatively consistent with the measured $\mathrm{P0}$--$\mathrm{P3}$ trend (Fig.~\ref{fig:size_effect}D). Along the confinement axis, $\chi$ likewise approaches a constant set by $u_0$ in the weak-confinement limit and is then amplified monotonically as confinement tightens (Fig.~\ref{fig:regulation}E), so confinement and polarity act cooperatively to enhance unequal partitioning.

\section*{Model captures embryonic morphology and lineage}
\begin{figure}[htbp]
\includegraphics[width=\textwidth]{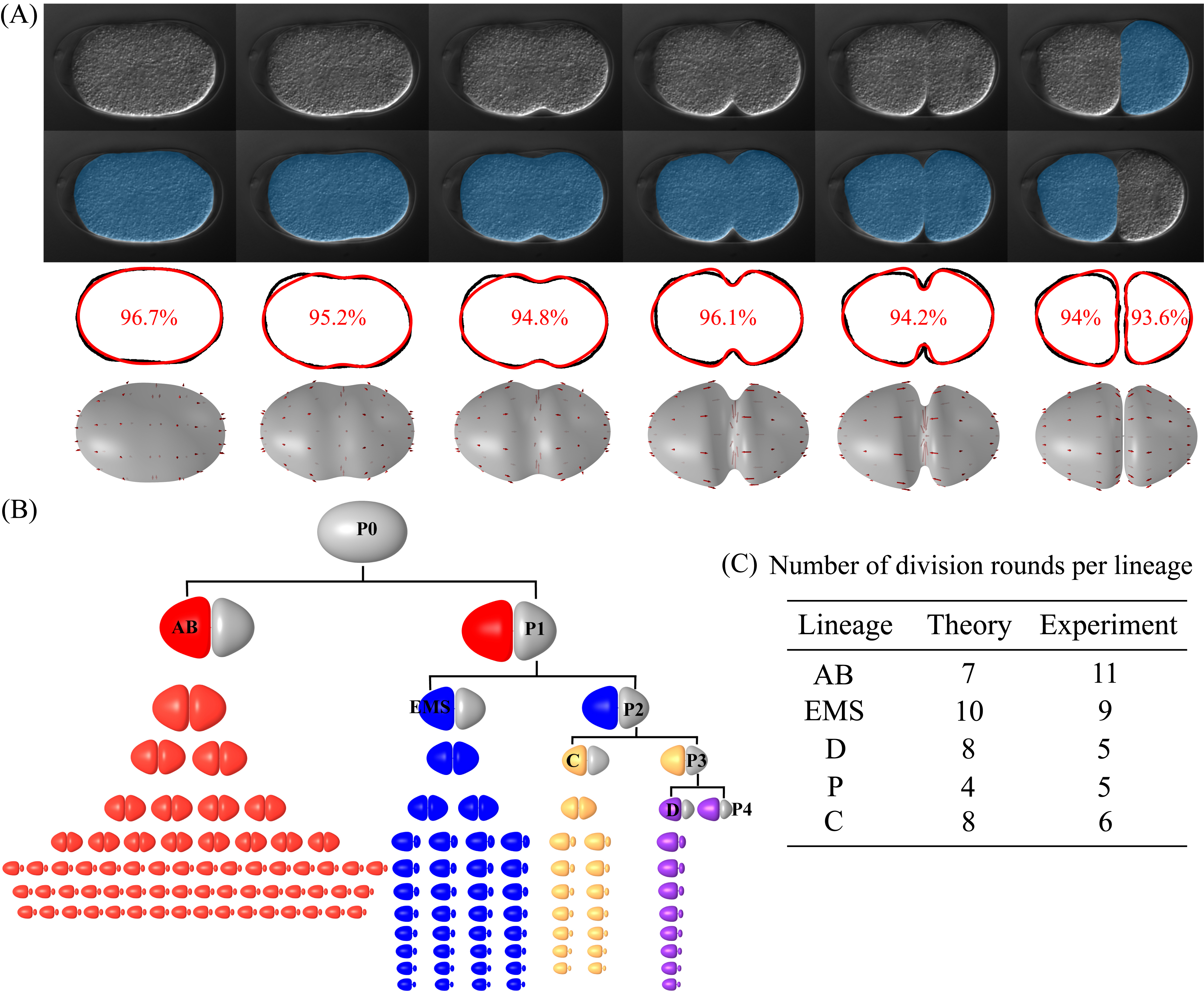}
\caption{\textbf{Active-surface model reproduces embryonic division morphology and captures the early lineage.} \textbf{A}, Simulated and experimental cortex shapes at six time points (columns) of the first (P0 cell) cleavage. Rows, top to bottom: brightfield images; the same images with the automatic segmentation~\cite{yamamoto2017,segmentation} overlaid in blue; axial cross-section contours (simulated in red, segmented experimental boundary in black) annotated with the similarity index (Supplementary Information); and the simulated three-dimensional cortex surface with red arrows showing the active force density. At the final time point the two daughters are shown separately in each row. \textbf{B}, Lineage-level simulation of early \textit{C.\,elegans} divisions from the P0 zygote, with a single shared parameter set (shell confinement $\epsilon=0.87$) and continuously decreasing cell size. Somatic blastomeres (AB, red; EMS, blue; C, orange; D, purple) divide without polarity regulation, and germline blastomeres (P0--P3, gray) divide with the localized $\kappa$-modulation of Fig.~\ref{fig:regulation}. A daughter smaller than the critical radius $\tilde{R}_{0,c}$ set by the pitchfork boundary of Fig.~\ref{fig:size_confinement} can no longer divide. \textbf{C}, Number of division rounds completed by each founder lineage, predicted by the model and measured \textit{in vivo}~\cite{Sulston1983,Guan2025}.}
\label{fig:validation}
\end{figure}

To validate the model against experiment, we compare simulated cortex contours with cell boundaries extracted from time-lapse imaging of the first (P0 cell) cleavage in \textit{C.\,elegans}~\cite{yamamoto2017,segmentation}. At six successive time points spanning the full constriction sequence—from the initial ellipsoidal configuration through progressive furrow deepening to near-complete division—the simulated contour (red) closely follows the experimental boundary (black), achieving a similarity index exceeding 90\% (Fig.~\ref{fig:validation}A; Supplementary Movie~4; defined in Methods). This quantitative agreement confirms that the model parameter set, with the shell confinement fixed at $\epsilon=0.87$, faithfully captures the cortex geometry throughout the division process.

We next ask whether the same framework can reproduce the complete early embryonic division pattern. Guided by the known \textit{C.\,elegans} lineage~\cite{Horvitz1977}, we let the somatic blastomeres (AB, EMS, C, and D) divide by self-organized mechanics alone and add the localized $\kappa$-modulation of Fig.~\ref{fig:regulation} to the successive germline divisions (P0--P3). All cells share a single parameter set (Supplementary Table~S3) at the same fixed shell confinement $\epsilon=0.87$, and only the cell size decreases during cleavage. The one calibrated input is the P0 zygote radius $\tilde{R}_0=5.4$, chosen to match the measured germline asymmetries of Fig.~\ref{fig:size_effect}D. From this starting point the model generates a lineage tree that captures the experimentally observed division pattern (Fig.~\ref{fig:validation}B). The AB lineage divides symmetrically while its cells remain large, producing an expanding array of equal-sized daughters. As the cells shrink, the size-induced instability eventually drives these somatic divisions asymmetric as well, a prediction we test against the AB lineage below (Fig.~\ref{fig:size_effect}C). The germline P blastomeres, by contrast, divide with progressively stronger asymmetry throughout, a prediction we compare directly against the measured germline series below (Fig.~\ref{fig:size_effect}D,E). The tree is built in a volume-driven manner. The two daughter volumes computed at each division become the inputs to the next round. The size at which divisions stop follows from the bifurcation analysis: it is the critical radius $\tilde{R}_{0,c}$ at which the fixed-activity trajectory crosses the pitchfork boundary $\tilde{\xi}_s=\tilde{\xi}_s^*(\tilde{R}_{0,c},\epsilon)$ into the failed-division regime (Fig.~\ref{fig:size_confinement}A,B, and Fig.~\ref{fig:regulation}D for the polarity-regulated germline). Below $\tilde{R}_{0,c}$ the mirror-symmetric branch is stable and the cortex relaxes to a weakly invaginated state, so the branch terminates. A generation whose cell number fails to double therefore marks an asymmetric division whose smaller daughter has dropped below $\tilde{R}_{0,c}$, and the lineage ends once all remaining cells lie in the failed-division regime. The model reproduces the finite number of divisions in every founder lineage, matching the measured count to within a single round for the EMS and P lineages, although lineage-specific deviations elsewhere show that size and confinement alone do not fix the exact number of rounds (Fig.~\ref{fig:validation}C). Taken together, these results show that the interplay between self-organized active-surface instability and localized polarity regulation captures the main features of the early division pattern, while additional control beyond geometry contributes to setting how many times each lineage divides.

\section*{Embryonic data support geometric control}
\begin{figure}[htbp]
\centering
\includegraphics[width=\textwidth]{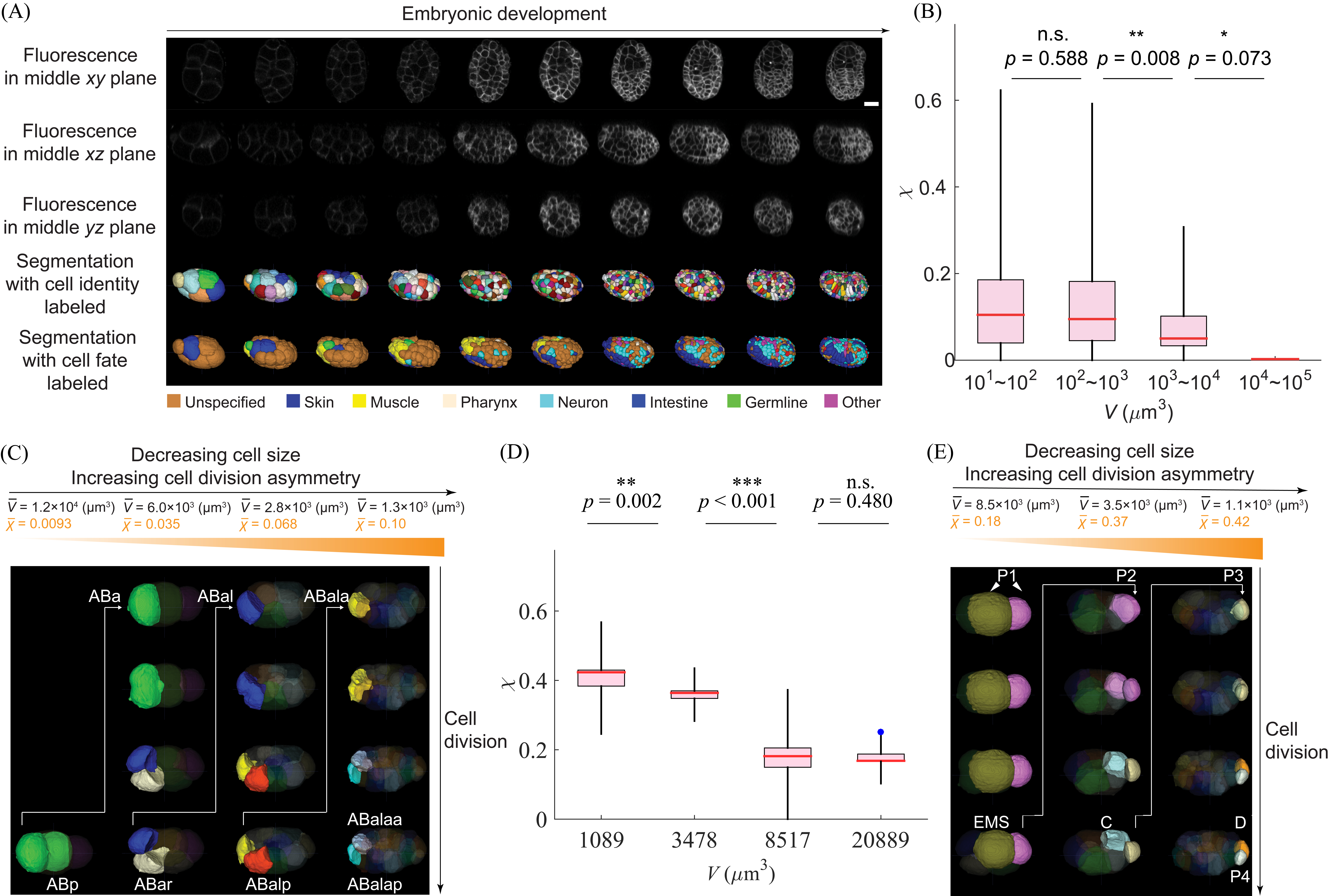}
\caption{\textbf{Cell-size-dependent division asymmetry across \textit{C.\,elegans} embryogenesis.} \textbf{A}, Time-lapse three-dimensional morphological maps of \textit{C.\,elegans} embryonic development from the \textit{CMap} dataset~\cite{Guan2025,cao2020}, showing fluorescence images in the middle $xy$, $xz$ and $yz$ planes (top three rows) together with cell segmentations colored by cell identity (fourth row) and cell fate (fifth row). \textbf{B}, Division asymmetry $\chi$ as a function of mother-cell volume $V$, pooled across all division events and binned into four volume decades ($10^1$--$10^2$, $10^2$--$10^3$, $10^3$--$10^4$, $10^4$--$10^5\,\mu\mathrm{m}^3$). Pairwise comparisons between adjacent bins are annotated (one-sided Wilcoxon rank-sum test; n.s., $p > 0.1$; $^{*}p < 0.1$; $^{**}p < 0.05$). In \textbf{B} and \textbf{D}, boxes span the first to third quartiles with the median as the central line, whiskers extend to the most extreme values within $3\,$IQR of the box, where IQR is the interquartile range, and points beyond are plotted as outliers (Supplementary Information). \textbf{C}, Representative snapshots along the AB somatic lineage (ABa, ABal, ABalap, ABalaa), with mean cell volume indicated above each panel and $\chi$ shown by the orange bar. \textbf{D}, $\chi$ as a function of mean cell volume $\bar{V}$ for the four successive P-lineage germline divisions (P0--P3; $\bar{V} = 2.089,\ 0.852,\ 0.348,\ 0.109 \times 10^4\,\mu\mathrm{m}^3$). Adjacent-stage comparisons: P0 versus P1, $p=0.480$ (n.s.); P1 versus P2, $p<0.001$ ($^{***}$); P2 versus P3, $p=0.002$ ($^{**}$) (one-sided Wilcoxon rank-sum test). \textbf{E}, Representative snapshots of the P-lineage germline blastomeres (P1, P2, P3) and their somatic daughters (EMS, C, D, P4).}
\label{fig:size_effect}
\end{figure}

To assess whether the mechanical trends identified in the model are operative in living embryos, we analyse two public four-dimensional morphological datasets of \textit{C.\,elegans} embryogenesis: \textit{CMap}, comprising 8 mechanically uncompressed embryos, and \textit{CShaper}, comprising 17 mechanically compressed embryos~\cite{Guan2025,cao2020}. Membrane and nuclear labels resolve every division and identify it by lineage across embryos (Fig.~\ref{fig:size_effect}A), so that each embryo contributes one value of $\chi$ per division event (Supplementary Information).

Pooling all division events from the \textit{CMap} dataset and binning by mother-cell volume reveals that $\chi$ increases systematically as cell size decreases (Fig.~\ref{fig:size_effect}B), in direct agreement with the model prediction of Fig.~\ref{fig:size_confinement}A. Representative examples from the AB somatic lineage (ABa $\to$ ABal $\to$ ABalap $\to$ ABalaa) show that division asymmetry escalates continuously as cells shrink through successive rounds of cleavage, confirming the prediction that even somatic divisions become asymmetric once the cells are sufficiently small (Fig.~\ref{fig:size_effect}C). The same holds for the four successive P-lineage divisions, where $\chi$ increases with decreasing mean cell volume as the geometric instability progressively dominates the polarity-specified positioning: the asymmetry is statistically indistinguishable between P0 and P1, but rises significantly at P2 and again at P3 (Fig.~\ref{fig:size_effect}D,E). The P0 size in the model was calibrated to reproduce this measured trend, so that the germline asymmetries computed along the model curve (arrows in Fig.~\ref{fig:regulation}D) tie the mechanical prediction directly to experiment. Both model and data show $\chi$ staying low across the early germline divisions and rising steeply for the late ones, although the model places this rise at P3, whereas the measured asymmetry is already pronounced at P2.

\begin{figure}[htbp]
\includegraphics[width=\linewidth,height=0.70\textheight,keepaspectratio]{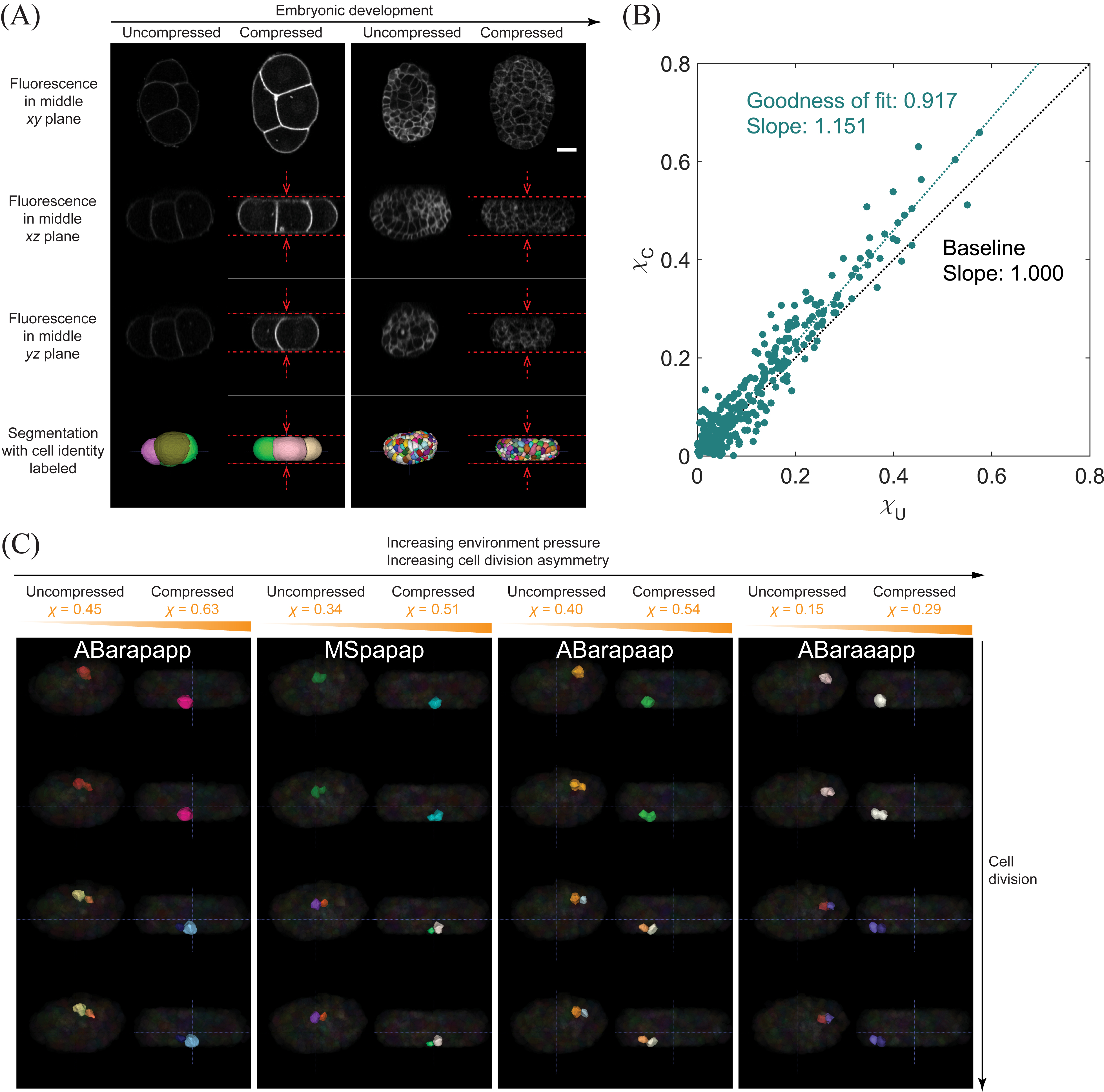}
\caption{\textbf{Mechanical confinement amplifies division asymmetry across \textit{C.\,elegans} embryogenesis.} \textbf{A}, Three-dimensional fluorescence imaging of uncompressed (\textit{CMap}) and compressed (\textit{CShaper}) \textit{C.\,elegans} embryos~\cite{cao2020,Guan2025} at an early developmental stage (left two columns, $\sim$4-cell) and a later stage (right two columns, many-cell), shown in the middle $xy$, $xz$ and $yz$ planes (top three rows) and as cell-identity-labeled segmentations (bottom row). \textbf{B}, Pairwise comparison of division asymmetry between all matched cell division events from uncompressed ($\chi_U$) and compressed ($\chi_C$) embryos. Each point is one unique cell division, the black dashed line is the identity (slope~$=1$), and linear regression (green dashed line) yields slope~$=1.151$ with $R^2=0.917$. \textbf{C}, Time-lapse images of four representative cell lineages (ABarapapp, MSpapap, ABarapaap, ABaraaapp) under uncompressed and compressed conditions, each at successive time points spanning cell division. $\chi$ is annotated above each pair, and the orange bar above each column indicates the direction of increasing division asymmetry.}
\label{fig:confinement_effect}
\end{figure}

We next test the prediction that external mechanical confinement amplifies division asymmetry. Comparing all matched division events between the uncompressed (\textit{CMap}, $\chi_U$) and compressed (\textit{CShaper}, $\chi_C$) datasets yields a tight proportional relationship, with a regression slope of $1.151$ ($R^2=0.917$), indicating that compression amplifies division asymmetry globally by $\sim$15\% relative to the uncompressed baseline (Fig.~\ref{fig:confinement_effect}B), consistent with the monotonic increase predicted by the model (Fig.~\ref{fig:size_confinement}E). The compressed embryo is visibly flattened relative to its uncompressed counterpart (Fig.~\ref{fig:confinement_effect}A), consistent with the reduced effective shell volume encoded in $\epsilon$, and individual lineages show the same compression-induced increase in $\chi$ (Fig.~\ref{fig:confinement_effect}C). Together, Figs.~\ref{fig:size_effect} and~\ref{fig:confinement_effect} indicate that the cell-size and confinement dependences identified in the model are operative \textit{in vivo}, and that the self-organized and polarity-regulated mechanisms cooperate to set the observed division asymmetry.

\section*{Implications and limitations}

Asymmetric cell division is a conserved strategy for generating cellular diversity, yet the physical principles that set the degree of asymmetry have remained elusive. Here we established that, beyond the molecular polarity cues usually believed responsible, asymmetric division can also be driven by a mechanical instability of the active cortex. This instability may complement and enhance the polarity-based patterning. Permitted by the broken up-down symmetry of the cortex~\cite{salbreux2017}, a curvature-dependent active stress makes symmetric division the default outcome, and either decreasing cell size or tightening confinement destabilizes that symmetry through a pitchfork bifurcation, so that two independent geometric quantities act on a single active-surface instability. Embryonic morphological data support both predicted control parameters.

Although curvature-dependent active stress is permitted by the broken up-down symmetry of the cortex~\cite{salbreux2017}, its microscopic origin remains an open question. Active elastic thin-shell theory offers one route~\cite{berthoumieux2014}: an active stress profile that varies across the finite cortical thickness generates, besides an active tension, an active bending moment, which in a curved shell produces an effective in-plane stress proportional to the local curvature (Methods). Within this picture the coupling coefficients $\xi_s$ and $\xi_b$ in Eq.~(\ref{eq:act}) reflect the gradient of myosin activity across the cortex, so that variations in cortical architecture that alter the transverse distribution of motor activity could directly tune the degree of division asymmetry.

The present model captures cortical mechanics at the cellular scale but does not resolve processes at the subcellular or supracellular level. Cell polarity enters only as a phenomenological modulation of bending rigidity, whereas the underlying anterior--posterior domain is maintained by reaction--diffusion--advection dynamics among PAR proteins on the membrane and in the cytosol~\cite{kemphues1988,sailer2015,tostevin2008,lim2021}. The mechanical influence of neighbouring blastomeres is captured by an idealized ellipsoidal shell rather than by explicit cell--cell contact forces~\cite{fickentscher2013,tian2020}, the broader context set by global embryonic constraints is not resolved~\cite{seirinlee2022,miao2022}, and the model assumes axial symmetry, whereas real \textit{C.~elegans} blastomeres are non-axisymmetric and subject to three-dimensional spatial constraints~\cite{ichbiah2023,vanbavel2023}. A natural path forward is to incorporate the curvature-dependent active stress identified here into a phase-field framework for multicellular morphogenesis~\cite{morphosim2023,kuang2022}, in which division timing, axis, and volume ratio could emerge self-consistently from the interplay of active cortical stress, cell--cell interactions, and geometric confinement.

\section*{Methods}
\subsection*{Active surface model}
We model the cortex as a closed, axisymmetric viscous surface $\mathbf{X}(u,\phi,t)$ (surface coordinates $u\in[0,1]$, $\phi\in[0,2\pi]$), characterized by its metric $g_{ij}$ and curvature tensor $C_{ij}$, with bending elasticity described by the Helfrich energy~\cite{helfrich1973,ouyang1987} and shape evolving with the cortical velocity field $\mathbf{v}$ as $d\mathbf{X}/dt=\mathbf{v}$~\cite{mietke2019self}. Full geometric conventions are given in the Supplementary Information.

The cortical flows are governed by the force balance equation
\begin{equation}
\mathrm{div}(\mathbf{T})=-\mathbf{f}^{\mathrm{ext}}\label{eq:forcebalance},
\end{equation}
where $\mathrm{div}(\cdot)=\mathbf{e}^i\cdot\partial_i(\cdot)$ is the surface divergence operator, $\mathbf{T}$ is the total surface stress tensor, and $\mathbf{f}^{\mathrm{ext}}$ comprises intracellular pressure, frictional drag with the cytoplasm, and normal forces from the confining shell. The intracellular pressure $p$ enters as a Lagrange multiplier that enforces conservation of the cell volume $V_\mathrm{cell}$ throughout the deformation. We solve for the instantaneous shape $\mathbf{X}$ and flow $\mathbf{v}$ using the variational framework developed in our previous work, in which stationarity of the Rayleigh dissipation functional yields the governing equations~\cite{Gao2026} (Supplementary Information).

The stress tensor $\mathbf{T}$ decomposes into equilibrium and dissipative contributions. The equilibrium part arises from the Helfrich bending energy $\mathcal{F}_H=\int dA\!\left[\gamma+\frac{\kappa}{2}(C^k_k)^2\right]$, which penalizes curvature deformations and sets the mechanical stiffness of the cortex. Varying $\mathcal{F}_H$ with respect to surface deformations yields the equilibrium in-plane and normal tension components
\begin{equation}
t^H_{ij}=\gamma g_{ij}+\frac{\kappa}{2}C^k_k\!\left(C^k_k g_{ij}-2C_{ij}\right), \qquad t^H_{i,n}=\kappa\nabla_i C^k_k,\label{eq:helfrich}
\end{equation}
where $\gamma$ is the isotropic surface tension and $\kappa$ the bending rigidity. The dissipative contributions arise from viscous cortical flows and from the active stresses generated by myosin motors~\cite{juli97,Prost2015naturephysics}, and the dissipative stress tensor reads
\begin{equation}
t^d_{ij}=2\eta_s\tilde{S}_{ij}+\eta_b\,\mathrm{div}(\mathbf{v})\,g_{ij}+t^a_{ij}\label{eq:dis},
\end{equation}
where $\eta_s$ and $\eta_b$ are the surface shear and bulk viscosities, $\tilde{S}_{ij}=S_{ij}-\tfrac{1}{2}S^k_k g_{ij}$ is the traceless strain-rate tensor, and $t^a_{ij}$ is the curvature-dependent active stress of Eq.~(\ref{eq:act}).

\subsection*{Relation to active thin-shell theory}
Modelling the cortex as an elastic shell of finite thickness $h$, an active stress profile $\zeta_a(z)$ that varies across the shell cross-section generates, upon integration through the thickness, both an active tension $t_a\propto\zeta_a(0)h$, proportional to the mean active stress, and an active bending moment $m_a\propto h^3\partial_z\zeta_a|_{z=0}$, proportional to the stress gradient~\cite{berthoumieux2014}. In a curved shell an active bending moment produces an effective in-plane stress proportional to the local curvature~\cite{berthoumieux2014}, yielding a curvature-dependent active tension at the two-dimensional level. A stronger transverse gradient, arising for instance from preferential binding of myosin to the inner face of the actin network, produces a larger active bending moment and hence stronger curvature coupling.

\subsection*{Nondimensionalization and control parameters}
We nondimensionalize the model parameters using the bending rigidity $\kappa_0$ as the energy scale and the hydrodynamic length $L_h=\sqrt{\eta_b/\Gamma}$, set by the bulk viscosity $\eta_b$ and the substrate friction $\Gamma$, as the length scale, with the corresponding time scale $\tau_\eta=\eta_b L_h^2/\kappa_0$. This yields the dimensionless active couplings $\tilde{\xi}_s=\xi_s\Delta\mu\,L_h/\kappa_0$ and $\tilde{\xi}_b=\xi_b\Delta\mu\,L_h/\kappa_0$. Cell size is measured by the reduced mother-cell radius $\tilde{R}_0=R_0/L_h$, where $R_0$ is defined by the cell volume $V_\mathrm{cell}=\tfrac{4\pi}{3}R_0^3$, and the degree of confinement by $\epsilon=V_\mathrm{cell}/V_\mathrm{shell}$, where $V_\mathrm{shell}$ is the volume of the ellipsoidal shell. The shared parameter set used throughout is listed in Supplementary Table~S3.

\subsection*{Numerical methods}
The time derivatives in the dynamic equations are discretized with an implicit finite-difference scheme, so that each time step reduces to a boundary value problem on $u\in[0,1]$. The resulting problems are solved with the MATLAB collocation solver \texttt{bvp5c}~\cite{bvpsolv01}, using relative and absolute tolerances of $10^{-5}$ and $10^{-8}$ and a maximum mesh size of $5\times10^4$. Stationary shapes are obtained by setting the time derivatives to zero and following the solution branches with numerical continuation in $\tilde{\xi}_s$ at fixed coupling ratio. Near a fold, where continuation in activity ceases to be single-valued, the total meridional arclength is used as the continuation coordinate and the activity is solved for as an additional unknown, which allows both sides of the turning point to be followed.

To probe the sensitivity of the outcome to the initial condition, a small random perturbation is applied to the active couplings at the first time step only: $\tilde{\xi}_s$ and $\tilde{\xi}_b$ are each augmented by $\sum_{n=1}^{10}c_n\cos(n\pi u)$, with the coefficients $c_n$ drawn independently from a uniform distribution on $[-10^{-2}\tilde{\xi},10^{-2}\tilde{\xi}]$, where $\tilde{\xi}$ denotes the coupling being perturbed. Each data point with error bars is obtained from ten independent realizations, and the error bars report the resulting standard deviation. The stability of a stationary state is classified dynamically: the state is taken as the initial condition, perturbed as above, and evolved for $40\,\tau_\eta$, and the state is stable when the distance between the time-dependent and stationary shapes decays and unstable when it grows (Supplementary Fig.~S5). A division is counted as complete once the furrow radius falls below the numerical cutoff $0.001\,R_0$. Full derivations, boundary conditions, and the definition of the distance function are given in the Supplementary Information.

\subsection*{Experimental data analysis and statistics}
We analysed two published cell-resolved four-dimensional morphological datasets of \textit{C.\,elegans} embryogenesis: \textit{CMap}, comprising eight mechanically uncompressed embryos, and \textit{CShaper}, comprising seventeen mechanically compressed embryos~\cite{Guan2025,cao2020}. For each cell identity the segmented volume is averaged over the cell lifespan, and the division asymmetry index $\chi=(V_b-V_s)/(V_b+V_s)$ is computed from the two daughter volumes, where $V_b$ and $V_s$ are the larger and smaller daughter. The statistical unit is therefore a single lineage-resolved division event in a single embryo, and every embryo contributes one value of $\chi$ per division event.

For the cell-size analysis, somatic cell division events from the uncompressed dataset are grouped into four mother-cell-volume intervals ($10^1$--$10^2$, $10^2$--$10^3$, $10^3$--$10^4$ and $10^4$--$10^5\,\mu\mathrm{m}^3$), and adjacent volume groups are compared with a one-sided Wilcoxon rank-sum test. Germline cell division events are compared between adjacent events with the same test. The one-sided form is used because the model predicts the direction of the effect in advance. Significance is annotated as n.s. for $p>0.1$, one asterisk for $p<0.1$, two for $p<0.05$ and three for $p<0.001$. In the box plots, the box spans the first to third quartiles, the central line marks the median, the whiskers extend to the most extreme values within $[Q_1-3\,\mathrm{IQR},Q_3+3\,\mathrm{IQR}]$, where $\mathrm{IQR}=Q_3-Q_1$ is the interquartile range, and values outside this range are plotted individually. For the confinement comparison, division events are matched between datasets by lineage identity and the matched values are fitted by linear regression, giving a slope of $1.151$ with $R^2=0.917$. Simulated and segmented cortex contours are compared through the similarity index $|A\cap B|/|A|$, where $A$ and $B$ are the sets of pixels enclosed by the experimental and simulated contours.

\section*{Data and code availability}
The data that support the findings of this study are available at \url{https://github.com/ruima86/AsymmetricCellDivision}. The code used to produce the simulation data in this paper is available from the project repository:\par\noindent \url{https://github.com/ruima86/AsymmetricCellDivision}.

\section*{Author contributions}
G.G., C.T., and R.M. designed research; D.G. and G.G. performed research; D.G., G.G., C.T., and R.M. analyzed data; and D.G., G.G., C.T., and R.M. wrote the paper.

\section*{Acknowledgments}
We gratefully acknowledge all members of Tang Lab and Ma Lab, with special thanks to Prof. Xiaojing Yang for insightful discussion. We thank Prof. Fan Jin for critical reading of our manuscript. This work was supported by the Fundamental Research Funds for the Central Universities of China under Grant No. 20720250069 to R.M. and the National Natural Science Foundation of China under Grant No. 12474199 to R.M., Grant No. 12090053 to C.T., and Grant No. 32088101 to C.T. and G.G. The authors used Claude (Anthropic, model Opus 5) to improve the language and readability of the manuscript. No figure, experimental image or measured data was generated or altered by AI software. The theoretical model, the simulation code, the data analysis and all scientific conclusions are the authors{\textquoteright} own. All content was reviewed and verified by the authors, who take full responsibility for it.

\bibliography{library}
\bibliographystyle{unsrt}

\clearpage
\setcounter{section}{0}
\setcounter{figure}{0}
\setcounter{table}{0}
\setcounter{equation}{0}
\renewcommand{\thefigure}{S\arabic{figure}}
\renewcommand{\thetable}{S\arabic{table}}
\renewcommand{\theequation}{S\arabic{equation}}

\begin{flushleft}
{\Large\textbf{Supplementary Information}}\\[0.4em]
{\large Cell size and confinement drive asymmetric cell division through a cortical instability}\\[0.3em]
Da Gao, Guoye Guan, Chao Tang and Rui Ma
\end{flushleft}

\vspace{0.5em}
\begingroup
\footnotesize
\setlength{\parskip}{0pt}
\tableofcontents
\endgroup
\newpage

\suppnote{Active-surface model}
We use the active-surface formulation introduced in ref.~\cite{Gao2026}, with the curvature-dependent active stress, confining shell, and localized polarity regulation specified below.

\subsection{Surface geometry and kinematics}

The geometry and force conventions are summarized in Supplementary Fig.~\ref{figS:Illustration}. We consider a deforming axisymmetric surface described by
\begin{equation}
\mathbf{X}(u, \phi, t)=[r(u,t)\cos\phi,r(u,t)\sin\phi, z(u, t)],
\end{equation}
where $r(u,t)$ and $z(u, t)$ denote time-dependent components of points on the surface, and $u \in [0, 1]$ and $\phi\in [0, 2\pi]$ are time-independent fixed mesh coordinates. The mapping between the mesh coordinate 
$u$ and the physical arc length $s$ on the deforming surface is given by the kinematic relation $s=h(t)u$. Additionally, the coordinate $r$ and $z$ depend on the geometric tangential angle $\psi$ through the scaling factor $h(t)$
\begin{align}
r' &= h \cos\psi\label{eq:dur} \\
z' &= -h \sin\psi\label{eq:duz}.
\end{align}
Given this, the tangent vectors $\mathbf{e}_i=\partial_i\mathbf{X}$ and surface normal $\mathbf{n}=\mathbf{e}_u\times\mathbf{e}_{\phi}/|\mathbf{e}_u\times\mathbf{e}_{\phi}|$ are given by
\begin{equation}\label{eq:bvs}
\mathbf{e}_u=h\begin{pmatrix}
\cos\psi\cos\phi \\
\cos\psi\sin\phi \\
-\sin\psi
\end{pmatrix}
\quad
\mathbf{e}_{\phi}=r\begin{pmatrix}
-\sin\phi\\
\cos\phi \\
0
\end{pmatrix}\\
\quad
\mathbf{n}=\begin{pmatrix}
\sin\psi\cos\phi \\
\sin\psi\sin\phi \\
\cos\psi
\end{pmatrix}.
\end{equation}
The metric tensor $g_{ij}=(\partial_i\mathbf{X})\cdot(\partial_j\mathbf{X})$, Levi-Civita
tensor $\epsilon_{ij}=\mathbf{n}\cdot(\mathbf{e}_i\times\mathbf{e}_j)$ as well as the curvature tensor \hbox{$C_{ij}=-\mathbf{n}\cdot\partial_i\partial_j\mathbf{X}$} are diagonal and read
\begin{align}
g_{ij}&=\begin{pmatrix}
g_{uu} & 0\\
0 & g_{\phi\phi}
\end{pmatrix}=\begin{pmatrix}
h^2 & 0\\
0 & r^2
\end{pmatrix}\label{eq:gijaxi}\\
\epsilon_{ij}&=\begin{pmatrix}
\epsilon_{uu} & \epsilon_{u\phi}\\
\epsilon_{{\phi}u} & \epsilon_{\phi\phi}
\end{pmatrix}=\sqrt{g}\begin{pmatrix}
0 & 1\\
-1 & 0
\end{pmatrix}\\
C_{ij}&=\begin{pmatrix}
C_{uu} & 0\\
0 & C_{\phi\phi}
\end{pmatrix}=
\begin{pmatrix}
h\psi'& 0\\
0 & r\sin\psi
\end{pmatrix}\label{eq:Cijaxi},
\end{align}
where $g=\text{det}(g_{ij})$.

The surface dynamics are described by the surface velocity field
\begin{equation}
\begin{split}
\mathbf{v}(u,\phi,t)
&= v^u(u,t)\,\mathbf{e}_u + v^\phi(u,t)\,\mathbf{e}_\phi + v_n(u,t)\,\mathbf{n} \\
&= \bar v_u(u,t)\,\bar{\mathbf{e}}_u + \bar v_\phi(u,t)\,\bar{\mathbf{e}}_\phi + v_n(u,t)\,\mathbf{n}.
\end{split}
\end{equation}
Following the arbitrary Lagrangian--Eulerian (ALE) framework~\cite{ALE2020}, the center-of-mass flow is $\mathbf{v}=d\mathbf{X}/dt$ with material derivative $d/dt=\partial_t+q^u\,\partial_u$, where the coordinate flow $q^u$ accounts for the motion of a surface point relative to the mesh coordinates. This yields
\begin{align}
\label{eq:meshqu}
q^u &= v^u - h^{-1}\!\left(\sin\psi\,\partial_t z - \cos\psi\,\partial_t r\right), \\
v_n &= \sin\psi\,\partial_t r + \cos\psi\,\partial_t z .
\end{align}

\subsection{Vector and tensor divergence}
The surface divergences of a vector field $\mathbf{v}$ and a tension tensor field $\mathbf{T}$ are
\begin{equation}
\text{div}(\mathbf{v})=\mathbf{e}^i\cdot\partial_i\mathbf{v}=\nabla_iv^i+C_k^{\,k}v_n\qquad\text{div}(\mathbf{T})=\mathbf{e}^i\cdot\partial_i\mathbf{T}=\nabla_i\mathbf{t}^i.
\end{equation}
Throughout, we use the normalized tangent basis vectors $\bar{\mathbf{e}}_u:=\mathbf{e}_u/h$ and $\bar{\mathbf{e}}_{\phi}:=\mathbf{e}_\phi/r$, with $\mathbf{e}_i$ given in Eqs.~(\ref{eq:bvs}), and decompose $\mathbf{v}=\bar{v}_i\bar{\mathbf{e}}_i+v_n\mathbf{n}$ and $\mathbf{T} = \bar{\mathrm{T}}_{ij}\bar{\mathbf{e}}_i\otimes\bar{\mathbf{e}}_j + \bar{\mathrm{T}}_{n,i}\bar{\mathbf{e}}_i\otimes\mathbf{n}$. The explicit component expressions for $\text{div}(\mathbf{v})$ and $\text{div}(\mathbf{T})$ on the axisymmetric surface are given in ref.~\cite{Gao2026} [Eqs.~(S13)--(S16) therein].
\subsection{Passive, viscous, and curvature-dependent active stresses}
The equilibrium tension and moment tensors arising from the Helfrich free energy $\mathcal{F}_H=\int dA\,[\gamma+\tfrac{\kappa}{2}(C^k_k)^2]$, the strain-rate tensor $\mathbf{S}=\tfrac12(\mathbf{e}_i\cdot\partial_j\mathbf{v}+\mathbf{e}_j\cdot\partial_i\mathbf{v})\,\mathbf{e}^i\otimes\mathbf{e}^j$ together with its isotropic and traceless parts $\mathbf{S}^{\mathrm{iso}}$ and $\tilde{\mathbf{S}}$, and the general force- and torque-balance equations on an axisymmetric surface, are all given in ref.~\cite{Gao2026} (Secs.~IV and~V therein) and are used here without modification; the present model has no spontaneous curvature ($C_0=0$). On the axisymmetric surface the curvature tensor has components $\bar{C}_{uu}=\psi'/h$ and $\bar{C}_{\phi\phi}=\sin\psi/r$, and its traceless part is
\begin{equation}
\tilde{\mathbf{C}}=\frac{1}{2}\left(\frac{\psi'}{h}-\frac{\sin\psi}{r}\right)\left(\bar{\mathbf{e}}_u\otimes\bar{\mathbf{e}}_u-\bar{\mathbf{e}}_\phi\otimes\bar{\mathbf{e}}_\phi\right),
\end{equation}
where the prefactor $\psi'/h-\sin\psi/r=C^u_u-C^\phi_\phi$ is the difference between the two principal curvatures.

The modification specific to the present work is the curvature-dependent active stress in the dissipative tension tensor $\mathbf{T}_d$. Instead of the constant isotropic active stress $\propto\xi\Delta\mu$ of ref.~\cite{Gao2026}, the active stress here is linear in the curvature tensor, with a deviatoric contribution $\propto\xi_s\Delta\mu\,\tilde{\mathbf{C}}$ and an isotropic contribution $\propto\xi_b\Delta\mu\,C^k_k\mathbf{G}$ that depends on the mean curvature. On the axisymmetric surface, $\mathbf{T}_d=(\bar{\mathrm{T}}_d)_{uu}\bar{\mathbf{e}}_u\otimes\bar{\mathbf{e}}_u+(\bar{\mathrm{T}}_d)_{\phi\phi}\bar{\mathbf{e}}_\phi\otimes\bar{\mathbf{e}}_\phi+(\bar{\mathrm{T}}_d)_{u\phi}(\bar{\mathbf{e}}_u\otimes\bar{\mathbf{e}}_\phi+\bar{\mathbf{e}}_\phi\otimes\bar{\mathbf{e}}_u)$, with components
\begin{align}
(\bar{\mathrm{T}}_d)_{uu} & = \eta_b(\bar{S}_{uu}+\bar{S}_{\phi\phi})+\eta_s(\bar{S}_{uu}-\bar{S}_{\phi\phi})+\frac{1}{2}\xi_s\Delta\mu\left(\bar{C}_{uu}-\bar{C}_{\phi\phi}\right)+\xi_b\Delta\mu(\bar{C}_{uu}+\bar{C}_{\phi\phi})\label{eq:Tuu}\\
(\bar{\mathrm{T}}_d)_{\phi\phi} & = \eta_b(\bar{S}_{uu}+\bar{S}_{\phi\phi})-\eta_s(\bar{S}_{uu}-\bar{S}_{\phi\phi})-\frac{1}{2}\xi_s\Delta\mu\left(\bar{C}_{uu}-\bar{C}_{\phi\phi}\right)+\xi_b\Delta\mu(\bar{C}_{uu}+\bar{C}_{\phi\phi})\label{eq:Tpp}\\
(\bar{\mathrm{T}}_d)_{u\phi} & =(\bar{T}_d)_{\phi u} = 2\eta_s \bar{S}_{u\phi}\label{eq:Tup},
\end{align}
where the strain-rate components in the normalized basis read~\cite{Gao2026}
\begin{align}
\bar{S}_{uu} &= \frac{1}{h}\bar{v}_u' + \frac{\psi'}{h}\,v_n,\label{eq:Suu}\\
\bar{S}_{\phi\phi} &= \frac{\cos\psi}{r}\,\bar{v}_u + \frac{\sin\psi}{r}\,v_n,\label{eq:Spp}\\
\bar{S}_{u\phi} &= \frac{1}{2}\left(\frac{\bar{v}_\phi'}{h} - \frac{\cos\psi}{r}\,\bar{v}_\phi\right),\label{eq:Sup}
\end{align}
with $\bar{v}_u,\bar{v}_\phi$ the tangential flow components in the normalized basis, $v_n$ the normal velocity, and a prime denoting $\partial_u$. The deviatoric active stress $\xi_s\Delta\mu\,\tilde{\mathbf{C}}$, proportional to the difference between the two principal curvatures, is the major contribution responsible for forming the strongly ingressed cleavage furrow.

The total tension tensor is $\mathbf{T}=\mathbf{T}_e+\mathbf{T}_d$, where $\mathbf{T}_d$ is the dissipative stress and $\mathbf{T}_e$ is the equilibrium (Helfrich) stress,
\begin{equation}
\label{eq:Te}
\mathbf{T}_e=t^H_{ij}\,\mathbf{e}^i\otimes\mathbf{e}^j+t^H_{i,n}\,\mathbf{e}^i\otimes\mathbf{n},
\end{equation}
whose components $t^H_{ij}$ and $t^H_{i,n}$ are given in Eq.~(\ref{eq:helfrich}). The force balance $\text{div}(\mathbf{T})=-\mathbf{f}^{\text{ext}}$ is given as Eq.~(\ref{eq:forcebalance}) in the main text.

\subsection{Confinement, volume constraint, and polarity regulation}
The cortex is enclosed by an axisymmetric ellipsoidal shell with major and minor semi-axes $a$ and $b$. Its volume is $V_{\mathrm{shell}}=4\pi a b^2/3$, and the degree of confinement is measured by
\begin{equation}
\epsilon=\frac{V_{\mathrm{cell}}}{V_{\mathrm{shell}}},
\end{equation}
where $V_{\mathrm{cell}}=4\pi R_0^3/3$ defines the equivalent mother-cell radius $R_0$. Conservation of $V_{\mathrm{cell}}$ is imposed by the pressure difference $p$, which is solved as a Lagrange multiplier. Contact with the confining shell is represented by the potential
\begin{equation}
P^{\mathrm{ext}}(u,\phi,t)=P_0\exp\!\left[\frac{z^2/a^2+r^2/b^2-1}{w_p^2}\right],
\label{con_po}
\end{equation}
with the corresponding normal force
\begin{equation}
\mathbf{f}^c=f_n^c\mathbf{n}
=-\left(\frac{\partial P^{\mathrm{ext}}}{\partial r}\sin\psi
+\frac{\partial P^{\mathrm{ext}}}{\partial z}\cos\psi\right)\mathbf{n}.
\label{eq:confforce}
\end{equation}

For polarity-regulated divisions, the bending rigidity is modulated by a localized Gaussian profile,
\begin{equation}
\kappa(u)=\kappa_0\left[1+\Delta_\kappa
\exp\!\left(-\frac{(u-u_0)^2}{\omega^2}\right)\right],
\end{equation}
where $u_0$, $\omega$, and $\Delta_\kappa$ specify the position, width, and amplitude of the cue. Somatic-lineage simulations use a spatially uniform bending rigidity, whereas the P-lineage simulations use the localized profile.

We nondimensionalize lengths by the hydrodynamic length $L_h=\sqrt{\eta_b/\Gamma}$, energies by $\kappa_0$, and times by $\tau_\eta=\eta_bL_h^2/\kappa_0$. The resulting control parameters include $\tilde R_0=R_0/L_h$, $\tilde\xi_s=\xi_s\Delta\mu L_h/\kappa_0$, and $\tilde\xi_b=\xi_b\Delta\mu L_h/\kappa_0$. The shared parameter set used for the lineage calculation is listed in Supplementary Table~\ref{tab:parameter}.

\suppnote{Variational formulation}
\subsection{Rayleigh dissipation functional}
We introduce the Rayleigh dissipation functional $\mathcal{R}$, whose stationary points
\begin{equation}
\label{eq:var}
\frac{\delta \mathcal{R}}{\delta \mathbf{v}} = 0
\end{equation}
have been shown to be equivalent to the force-balance equation, Eq.~(\ref{eq:forcebalance}) in the main text (see Ref.~\cite{Gao2026} for a detailed derivation).

Within the Onsager framework, the Rayleighian is given by the sum of the rate of change of the free energy and the entropy production rate,
\begin{equation}
\label{eq:rayleigh_Simple}
\mathcal{R}(\mathbf{v}, r_p)
=
\frac{dF}{dt}
+
\mathcal{D},
\end{equation}
where $dF/dt$ represents the reversible contribution associated with conservative forces, and $\mathcal{D}$ denotes the dissipation function, which is quadratic in the generalized velocities and encodes irreversible processes. For active surfaces, $\mathcal{D}$ also includes contributions from nonequilibrium driving, such as chemical activity.

In the following, we provide explicit expressions for the free energy change rate $dF/dt$ and the entropy production rate $\mathcal{D}$.

\subsection{Free energy change rate}
The equilibrium elastic properties of the surface are described by the Helfrich free-energy density $f_H=\gamma_H+f_\kappa$~\cite{helfrich1973}, with passive surface tension $\gamma_H$ and bending energy density
\begin{equation}
\label{eq:EnergyD_Bending}
f_\kappa=\frac{1}{2}\kappa\left(C^k_{\;k}\right)^2=\frac{1}{2}\kappa\left(\frac{\psi'}{h}+\frac{\sin\psi}{r}\right)^2
\end{equation}
(the Gaussian-curvature term is omitted, as the surface has spherical topology). The surface is additionally driven out of equilibrium by chemical reactions with chemical free-energy density $f_c(c_f,c_p)$, whose chemical potentials $\mu_n=\partial f_c/\partial c_n$ ($n=f,p$)---corresponding, in biological systems, to the reactants and products of ATP hydrolysis~\cite{juli97,Prost2015naturephysics}---define the chemical potential difference $\Delta\mu=\mu_f-\mu_p$ that acts as the nonequilibrium driving force generating the active stresses. Finally, a velocity-independent external force $\mathbf{f}_e^{\mathrm{ext}}=p\mathbf{n}+f_n^c\mathbf{n}$ arises from the intracellular pressure $p$ and from the normal force $f_n^c$ exerted by the confining potential [Eq.~(\ref{eq:confforce})].

Applying the surface transport theorem [ref.~\cite{Gao2026}, Eq.~(S36) therein] to the elastic, chemical, and external-force contributions yields the total free-energy change rate
\begin{equation}
\label{eq:dfdt}
\frac{dF}{dt}
=
2\pi \int_0^1
\left[
\partial_t f_\kappa
+ \frac{f_\kappa}{r h}\,\partial_t(rh)
+ \gamma\,\mathrm{div}(\mathbf{v})
- r_p\,\Delta\mu
+ J_n^{\mathrm{ext}}- \mathbf{f}^{\mathrm{ext}}_{e}\cdot\mathbf{v}
\right]
r h\, du +2\pi\left.(f_\kappa q^urh)\right|_0^1,
\end{equation}
where $\gamma=\gamma_e+\gamma_H$ is the total passive tension with effective chemical tension $\gamma_e=f_c-\mu_f c_f-\mu_p c_p$, $r_p=-r_f$ is the production rate of the product species, and $J_n^{\mathrm{ext}}=J_{n,f}\mu_f+J_{n,p}\mu_p$ is the influx of chemical free energy from the environment. We assume $J_n^{\text{ext}}$ is such that $\Delta\mu$ is maintained constant in space and time.

\subsection{Entropy production}
The dissipative part of the Rayleigh functional $\mathcal{D}$ has contributions from the internal and external dissipation, respectively. The internal dissipation is defined as
\begin{equation}\label{eq:entrprod_old}
\Theta_{\text{int}}=\int dA\left(\mathbf{T}_d :{\bf S}+r_p\Delta\mu\right).
\end{equation}
Note that thermodynamic variables ($\mathbf{T}_d$,$\Delta\mu$) are even, while their conjugate partners ($\mathbf{S}$,$r_p$) are odd under time-reversal. The $\Delta\mu$ is commonly treated as the thermodynamic force~\cite{juli97}.  Given this, the thermodynamic flux $\mathbf{T}_d$ is expressed as the constitutive law
\begin{equation}\label{eq:consti_Td}
\mathbf{T}_d = 2\eta_s\tilde{\mathbf{S}} + \eta_b\text{div}(\mathbf{v})\mathbf{G} + \xi_s\Delta\mu\tilde{\mathbf{C}}+\xi_b \Delta\mu C_k^k\mathbf{G},  
\end{equation}
as presented in Eq.~\ref{eq:dis} and the thermodynamic flux $r_p$ is expressed as
\begin{equation}\label{eq:rp_reactive}
\begin{split}
r_p&=-\xi_s\tilde{\mathbf{C}}:\tilde{\mathbf{S}}-2\xi_b {\mathbf{C}}^{\text{iso}}:\mathbf{S}^{\text{iso}}+\Lambda\Delta\mu\\
&=-\xi_s\tilde{\mathbf{C}}:\tilde{\mathbf{S}}-\xi_bC^k_k\text{div}(\mathbf{v})+\Lambda\Delta\mu
\end{split}
\end{equation}
with the phenomenological Onsager coefficients $\eta_s,\eta_b,\Lambda,\xi_s$ and $\xi_b$. Onsager reciprocity relations require that the reactive coefficients of cross-coupling must adopt opposite signs with $\xi_s$ ($\xi_b$) and $-\xi_s$ ($-\xi_b$) in the constitutive laws Eqs.~(\ref{eq:consti_Td}) and (\ref{eq:rp_reactive})~\cite{julicher2018}.
Note that the reactive part $\xi_s$ and $\xi_b$ do not contribute to entropy production and dissipation in non-equilibrium systems. Consequently, it is necessary to introduce a reformulated expression for entropy production that depends exclusively on the dissipative Onsager coefficients. In this context, the chemical production rate $r_p$ is adopted as the thermodynamic force instead of $\Delta\mu$
\begin{equation}
    \Delta\hat{\mu} =\frac{\xi_s}{\Lambda}\tilde{\mathbf{C}}:\tilde{\mathbf{S}}+\frac{\xi_b}{\Lambda}C^k_k\text{div}(\mathbf{v})+\frac{1}{\Lambda}r_p,\label{eq:CR2}
\end{equation}
 the constitutive law corresponding to the thermodynamic flux $\mathbf{T}_d$ reads
\begin{equation}
\begin{split}
\label{eq:CR1}
\hat{\mathbf{T}}_d &= 2\eta_s\tilde{\mathbf{S}} + \left(\eta_b+\frac{\xi_b^2}{\Lambda}\left(C_k^k\right)^2\right)\text{div}(\mathbf{v})\mathbf{G} + \frac{\xi_s\xi_b}{\Lambda}C_k^k\text{div}(\mathbf{v})\tilde{\mathbf{C}}+\frac{{\xi_s}^2}{\Lambda}(\tilde{\mathbf{C}}:\tilde{\mathbf{S}})\tilde{\mathbf{C}}\\
&+\frac{\xi_b\xi_s}{\Lambda}C_k^k(\tilde{\mathbf{C}}:\tilde{\mathbf{S}})\mathbf{G}
+\frac{\xi_s}{\Lambda}r_p \tilde{\mathbf{C}}+\frac{\xi_b}{\Lambda}C_k^k r_p\mathbf{G}.
\end{split}
\end{equation}
The Onsager coefficients~$\xi_s/\Lambda$ and $\xi_b C_k^k/\Lambda$ now appear as dissipative terms with the same sign in both constitutive laws Eqs.~(\ref{eq:CR2}) and (\ref{eq:CR1}) and cross-couples flux-force pairs $(\hat{\mathbf{T}}_d,r_p)$ with opposite time-reversal signature.
The new internal entropy production $\hat{\Theta}_{\text{int}}$ associated with constitutive laws Eqs.~(\ref{eq:CR2}) and (\ref{eq:CR1}) is written as
\begin{equation}\label{eq:Entropy_Dis}
\begin{split}
\hat{\Theta}_{\text{int}}(\mathbf{v},r_p)&=\int\left(\hat{\mathbf{T}}_d :\mathbf{S}+r_p\Delta\hat{\mu}\right) dA\\
&=\int(S_1+S_2+S_3)dA,
\end{split}
\end{equation}
where 
\begin{equation}\label{eq:InterDise}
\begin{split}
S_1 & = \eta_b \left(\bar{S}_{uu}+\bar{S}_{\phi\phi}\right)^2\\
S_2 & = \eta_s \left[\left(\bar{S}_{uu}-\bar{S}_{\phi\phi}\right)^2+\left(2\bar{S}_{u\phi}\right)^2\right]\\
S_3 & = \frac{1}{\Lambda}\left[r_p+\xi_b\left(\bar{C}_{uu}+\bar{C}_{\phi\phi}\right)\left(\bar{S}_{uu}+\bar{S}_{\phi\phi}\right)+\frac{\xi_s}{2}\left(\bar{C}_{uu}-\bar{C}_{\phi\phi}\right)\left(\bar{S}_{uu}-\bar{S}_{\phi\phi}\right)\right]^2.
\end{split}
\end{equation}
The external frictional forces introduced in the force-balance equation [Eq.~(\ref{eq:forcebalance}) in the main text] will additionally contribute to entropy production in the form
\begin{equation}\label{eq:entrprodext}
\Theta_{\text{ext}}=-\int\mathbf{v}\cdot\mathbf{f}^{\text{ext}}_d dA= 2\pi\int_0^1\left[\Gamma\bar{v}_u^2+\Gamma\bar{v}_\phi^2\right]rhdu
\end{equation}
where $\mathbf{f}_d^{\text{ext}}=-\Gamma\mathbf{v}_\parallel$ is the tangential frictional force, with $\mathbf{v}_\parallel=\bar v_u\bar{\mathbf{e}}_u+\bar v_\phi\bar{\mathbf{e}}_\phi$ the in-plane surface velocity.
Taking everything into account, the dissipative part of the Rayleigh functional $\mathcal{D}$ is then given by
\begin{equation}\label{eq:disspart}
\mathcal{D}(\mathbf{v},r_p)=\frac{1}{2}\left(\hat{\Theta}_{\text{int}}+\Theta_{\text{ext}}\right).
\end{equation}

\subsection{Enforcing geometric relations via Lagrange multipliers}
To enforce the SLE gauge condition $h'=0$, as well as the geometric relations that relate $r$ and $z$ to $h$ and $\psi$, we introduce Lagrange multipliers following the standard approach~\cite{seif91}.
The Rayleigh dissipation functional, therefore, acquires additional constraint terms.
\begin{equation}
\label{eq:Lmultiplier}
\mathcal{L}
=
2\pi\int_0^1 du\,
\partial_t\!\left[
\alpha\left(r'-h\cos\psi\right)
+\beta\left(z'+h\sin\psi\right)
+\zeta\, h'
\right],
\end{equation}
where $\alpha$, $\beta$, and $\zeta$ are Lagrange multipliers enforcing the
corresponding geometric constraints.

Collecting all contributions, the final constrained Rayleigh dissipation
functional reads
\begin{equation}
\label{eq:FinalRay}
\begin{split}
\bar{\mathcal{R}}
&= \mathcal{R} + \mathcal{L} \\
&=
2\pi\int_0^1
\Bigg[
\partial_t f_\kappa
+ f_\kappa\,\frac{\partial_t(rh)}{rh}
- r_p\,\Delta\mu
- \mathbf{f}_e^{\mathrm{ext}}\!\cdot\mathbf{v}
+ \gamma\,\mathrm{div}(\mathbf{v})+ \frac{1}{2}\left(S_1+S_2+S_3\right)+\frac{1}{2}\Gamma\left(\bar{v}_u^2+\bar{v}_\phi^2\right)\\
&+ J_n^{\mathrm{ext}}
\Bigg]
r h\, du+ 2\pi\int_0^1 du\,
\partial_t\!\left[
\alpha\left(r'-h\cos\psi\right)
+\beta\left(z'+h\sin\psi\right)
+\zeta\, h'
\right]+2\pi\left.(f_\kappa q^urh)\right|_0^1.
\end{split}
\end{equation}
Here $\mathcal{R}$ and $\mathcal{L}$ are defined in
Eqs.~\eqref{eq:rayleigh_Simple} and \eqref{eq:Lmultiplier}, respectively. The velocity-independent external force is purely normal, $\mathbf{f}_e^{\text{ext}}=f_n^{\text{ext}}\mathbf{n}$, and enters the Rayleighian only through the term $-\mathbf{f}_e^{\text{ext}}\cdot\mathbf{v}=-f_n^{\text{ext}}v_n$ in Eq.~(\ref{eq:FinalRay}). The tangential frictional force is not part of $\mathbf{f}_e^{\text{ext}}$ and is instead represented by the dissipation term $\tfrac{1}{2}\Gamma(\bar{v}_u^2+\bar{v}_\phi^2)$. Specific to the present work is the normal external force, with
\begin{equation}
f_n^{\text{ext}}= p + f_n^c,
\end{equation}
which includes the pressure difference $p$ that enforces conservation of the enclosed volume and the confinement force $f_n^c$ defined from Eq.~(\ref{con_po}).

In the following, the constrained Rayleigh dissipation functional
$\bar{\mathcal{R}}$ is varied with respect to the generalized variables
\[
\Phi \in
\left\{
v^u,\,
v^\phi,\,
\partial_t r,\,
\partial_t z,\,
\partial_t h,\,
\partial_t \psi,\,
\partial_t \alpha,\\,
\partial_t \beta,\,
\partial_t \zeta
\right\},
\]
by imposing the stationarity condition
\begin{equation}
\label{eq:Funcvar}
\frac{\delta \bar{\mathcal{R}}}{\delta \Phi} = 0 .
\end{equation}

\subsection{Force balance from stationarity}\label{Derived_ODE}
The density of the unconstrained Rayleighian $\mathcal{R}(\mathbf{v},r_p)$ [Eq.~(\ref{eq:rayleigh_Simple})] reads
\begin{equation}
\begin{split}\label{eq:RayDe}
R(\mathbf{v},r_p) &= \bigg[\partial_t f_\kappa+f_\kappa\frac{\partial_t(r h)}{rh} - r_p \Delta\mu + J_n^{\mathrm{ext}}-\mathbf{f}^{\text{ext}}_e\cdot\mathbf{v}+\gamma\text{div}(\mathbf{v})\\
&+\frac{1}{2}(S_1+S_2+S_3)+\frac{1}{2}\Gamma\left(\bar{v}_{u}^2+\bar{v}_\phi^2\right)\bigg]rh,
\end{split}
\end{equation}
with $f_\kappa$ given in Eq.~(\ref{eq:EnergyD_Bending}) and $(S_1+S_2+S_3)$ given in Eq.~(\ref{eq:InterDise}). The density of the constrained functional $\bar{\mathcal{R}}$ [Eq.~(\ref{eq:FinalRay})] additionally includes the Lagrange-multiplier terms,
\begin{equation}\label{eq:Rbaraxi}
\bar{R}=R(\mathbf{v},r_p)+\partial_t\left[\alpha(r'-h \cos\psi)+\beta(z'+h\sin\psi)+\zeta h'\right].
\end{equation}
Following the same variational procedure as ref.~\cite{Gao2026}, the stationarity conditions $\delta\bar{\mathcal{R}}/\delta\Phi=0$ [Eq.~(\ref{eq:Funcvar})] reproduce the force-balance equations of the active surface; we summarize the resulting equations below and refer to ref.~\cite{Gao2026} for the detailed algebra.

The variation with respect to the reaction rate, $\delta\bar{\mathcal{R}}/\delta r_p=0$, gives
 \begin{equation}\label{eq:rpconst}
     \frac{1}{\Lambda}\left[r_p+\xi_b \left(\bar{C}_{uu}+\bar{C}_{\phi\phi}\right)\left(\bar{S}_{uu}+\bar{S}_{\phi\phi}\right)+\frac{\xi_s}{2} \left(\bar{C}_{uu}-\bar{C}_{\phi\phi}\right)\left(\bar{S}_{uu}-\bar{S}_{\phi\phi}\right)\right]-\Delta \mu = 0,
 \end{equation}
from which the constitutive law for the production rate $r_p$ [Eq.~(\ref{eq:CR2})] follows. Variations with respect to the tangential flows $\bar{v}_u$ and $\bar{v}_\phi$ reproduce the meridional and azimuthal components of the tangential force balance. The production rate $r_p$ arising from these variations is replaced by $\Delta\mu$ through Eq.~(\ref{eq:rpconst}), so that the dissipative stress enters in its $\Delta\mu$ form $\mathbf{T}_d$,
\begin{align}
\bar{{\bf e}}_u \cdot {\rm div}(\mathbf{T}_d +\mathbf{T}_e)&=\Gamma\bar{v}_u-f^\text{ext}_u,\\
\bar{{\bf e}}_\phi \cdot {\rm div}(\mathbf{T}_d +\mathbf{T}_e)&=\Gamma\bar{v}_\phi-f^\text{ext}_\phi.
\end{align}
Variations with respect to $\partial_t r$, $\partial_t z$, $\partial_t\psi$ and $\partial_t h$, together with the geometric relations imposed through the Lagrange multipliers $\alpha$, $\beta$ and $\zeta$, reproduce the normal force balance. As in ref.~\cite{Gao2026}, the boundary term of the $h$ variation enforces $\zeta(0)=\zeta(1)=0$, which together with $\zeta''=0$ yields $\zeta=0$; eliminating $\alpha$ and $\beta$ then leads to
\begin{equation}
{\bf n} \cdot {\rm div}(\mathbf{T}_d+{\bf T}_e)  = -f_n^{\rm{ext}}.
\end{equation}
As in the tangential components, the production rate $r_p$ arising from the variation is replaced by $\Delta\mu$ through Eq.~(\ref{eq:rpconst}), so that all three force-balance equations use the same dissipative stress $\mathbf{T}_d$.
In these variational force-balance equations, the dissipative stress $\mathbf{T}_d$ and the friction force $\Gamma\mathbf{v}$ arise from variations of the dissipation rate $\mathcal{D}$, while the equilibrium stress $\mathbf{T}_e$ and the velocity-independent external force $\mathbf{f}^{\text{ext}}$ arise from variations of the free-energy change rate $dF/dt$.

\suppnote{Numerical methods and bifurcation analysis}
\subsection{Boundary conditions from variations}\label{Appendix:boundary}
The variations Eq.~(\ref{eq:Funcvar}) generate the boundary conditions including
\begin{align}
\left[\frac{\partial\bar{R}}{\partial[(\bar{v}_u)']}+\frac{\partial (f_\kappa q^u r h)} {\partial \bar{v}_u}\right]\delta \bar{v}_u\bigg\vert _{u=0}^{u=1}&= r\bar{\mathbf{e}}_u\cdot\left[(f_\kappa+\gamma) \mathbf{G}+\mathbf{T}_d\right]\cdot\bar{\mathbf{e}}_u\delta \bar{v}_u\bigg\vert_{u=0}^{u=1}\label{eq:Boundary_vu}\\
\frac{\partial\bar{R}}{\partial[(\bar{v}_{\phi})']}\delta \bar{v}_{\phi}\bigg\vert _{u=0}^{u=1} &= r\bar{\mathbf{e}}_u\cdot\left[(f_\kappa+\gamma) \mathbf{G}+\mathbf{T}_d\right]\cdot\bar{\mathbf{e}}_{\phi}\delta \bar{v}_{\phi}\bigg\vert_{u=0}^{u=1}\label{eq:Boundary_vp}\\
\left[\frac{\partial\bar{R}}{\partial[(\partial_t r)']}+\frac{\partial (f_\kappa q^u r h)} {\partial(\partial_t r)}\right]\delta (\partial_t r)\bigg\vert _{u=0}^{u=1} &= (\alpha-f_\kappa r\cos\psi) \delta (\partial_t r)\bigg\vert_{u=0}^{u=1}\label{Boundary_alpha}\\
\left[\frac{\partial\bar{R}}{\partial[(\partial_t z)']}+\frac{\partial(f_\kappa q^u r h)} {\partial (\partial_t z)} \right]\delta (\partial_t z)\bigg\vert _{u=0}^{u=1} &=  (\beta+f_\kappa r\sin\psi) \delta (\partial_t z)\bigg\vert_{u=0}^{u=1}\label{Boundary_beta}\\
\frac{\partial \bar{R}}{\partial[(\partial_t \psi)']}\delta (\partial_t \psi)\bigg\vert _{u=0}^{u=1} &= r\,\bar{\mathbf{e}}_u\cdot\mathbf{M}_e\cdot\mathbf{\bar{e}}_\phi \delta (\partial_t \psi)\bigg\vert_{u=0}^{u=1}\label{eq:Boundary_psi},
\end{align}
where the equilibrium moment $\mathbf{M}_e=(\bar{M}_e)_{ij}\bar{\mathbf{e}}_i\otimes\bar{\mathbf{e}}_j$ , the metric tensor $\mathbf{G}=\bar{\mathbf{e}}_u\otimes\bar{\mathbf{e}}_u+\bar{\mathbf{e}}_\phi\otimes\bar{\mathbf{e}}_\phi$, and the dissipative tension $\mathbf{T}_d=(\bar{\mathrm{T}}_d)_{ij}\bar{\mathbf{e}}_i\otimes\bar{\mathbf{e}}_j$. The boundary conditions are summarized in Tab.~\ref{Table:Boundry}.

\subsection{Formulation of boundary value problems}
In the ODE system summarized in Tab.~\ref{table:ODE}, all equations are formulated as functions of the meridional coordinate $u$. To compute time-dependent solutions, the time derivatives $\partial_t r$, $\partial_t z$, and $\partial_t \psi$ appearing in these equations are discretized using a finite difference scheme, thereby eliminating explicit derivatives with respect to $t$. After this temporal discretization, the governing equations reduce to a
boundary value problem (BVP) defined on the fixed interval $u\in[0,1]$.
For the dynamic equations, we set
\begin{equation}
\begin{split}
    \partial_t r & = \frac{r(u,t)-r(u,t-\Delta t)}{\Delta t}\\
    \partial_t z & = \frac{z(u,t)-z(u,t-\Delta t)}{\Delta t}\\
    \partial_t \psi & = \frac{\psi(u,t)-\psi(u,t-\Delta t)}{\Delta t},\\
\end{split}
\end{equation}
and using $r(u,t-\Delta t), z(u,t-\Delta t), \psi(u,t-\Delta t)$ from the previous time-step. This represents an implicit scheme to numerically integrate the equations. Solutions $[r(u,t),z(u,t),\psi(u,t),v_u,v_{\phi}]$ of the ODE system derived from the variational formulation then directly correspond to the surface geometries and flows at the next time point. 
Tab.~\ref{table:ODE} provides nine equations that are equivalent to twelve first-order ODEs. To enforce volume conservation, we introduce the auxiliary volume function $V(u)$, which
satisfies
\begin{equation}
\label{eq:volume}
    V' = \pi r^2 h \sin\psi ,
\end{equation}
together with the boundary conditions
\begin{equation}
\label{eq:bc_volume}
    V(0) = 0, \qquad
    V(1) = V_{\mathrm{cell}} = \frac{4\pi}{3} R_0^3 .
\end{equation}
Combining Eq.~\eqref{eq:volume} with the ODEs listed in
Tab.~\ref{table:ODE} yields a total of thirteen first-order ODEs.
Tab.~\ref{Table:Boundry} specifies twelve boundary conditions; together with Eq.~\eqref{eq:bc_volume}, there are fourteen boundary conditions in total. Including the additional unknown Lagrange multiplier $p$ that enforces volume conservation, these ODEs and boundary conditions constitute a well-posed BVP. We solve this problem using standard BVP solvers implemented in MATLAB~\cite{bvpsolv01}.

\subsection{Numerical continuation and stability classification}
\begin{figure}[H]
\centering
\includegraphics[width=0.92\textwidth]{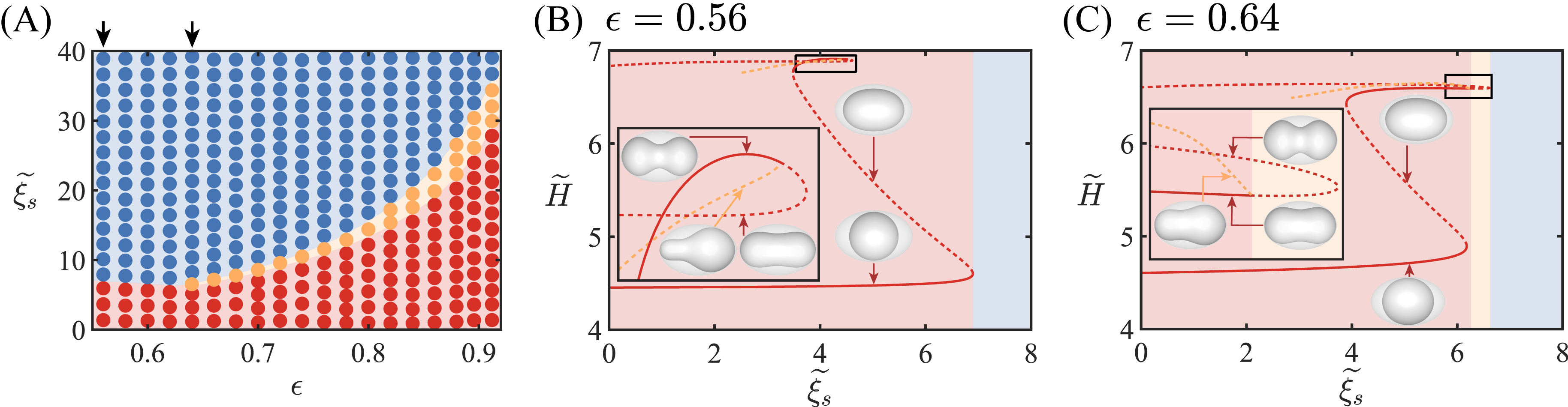}
\caption{\textbf{Bistability and subcritical bifurcation at weak confinement.}
\textbf{(A)} Phase diagram in the $(\epsilon, \tilde{\xi}_s)$ parameter plane, with the same regime colors as Fig.~\ref{fig:size_confinement} (blue: symmetric division; orange: asymmetric division; red: failed division). Arrows indicate the two confinement values examined in panels \textbf{B} and \textbf{C}.
\textbf{(B,C)} Bifurcation landscapes $\tilde{H}$ vs.\ $\tilde{\xi}_s$ at $\epsilon = 0.56$ (\textbf{B}) and $\epsilon = 0.64$ (\textbf{C}). Solid lines: stable branches; dashed lines: unstable branches. In both cases the symmetric branch is S-shaped, with two folding points, one at small $\tilde{H}$ and one at large $\tilde{H}$, whose ordering along $\tilde{\xi}_s$ sets the outcome as $\tilde{\xi}_s$ increases. At $\epsilon = 0.64$ (\textbf{C}) the large-$\tilde{H}$ fold lies at the higher $\tilde{\xi}_s$, so the cortex leaves the symmetric branch through a symmetry-breaking bifurcation and passes through an asymmetric-division window before symmetric constriction. At $\epsilon = 0.56$ (\textbf{B}) the small-$\tilde{H}$ fold lies at the higher $\tilde{\xi}_s$, so the cortex remains symmetric until that fold and then passes directly to symmetric constriction, with no asymmetric-division window. Arrows indicate the transitions between branches, and insets show representative cortex shapes at selected points.}
\label{figS:subcritical}
\end{figure}

Stationary shapes are obtained by setting the time derivatives in the dynamic equations to zero and solving the resulting BVP. We first construct a passive reference shape at zero activity and then follow the solution as the active couplings are varied, using each converged solution as the initial guess for the neighboring parameter value. The ratio between the two active couplings is held fixed while $\tilde\xi_s$ is used as the activity coordinate. Near a fold, where direct activity continuation ceases to be single-valued, the total meridional arclength is prescribed as the continuation coordinate and the active strength is solved as an additional unknown. This procedure allows both sides of the turning point to be followed and produces the stationary branches shown in Fig.~\ref{fig:size_confinement} and Supplementary Fig.~\ref{figS:subcritical}.

\begin{figure}[H]
\centering
\includegraphics[width=0.92\textwidth]{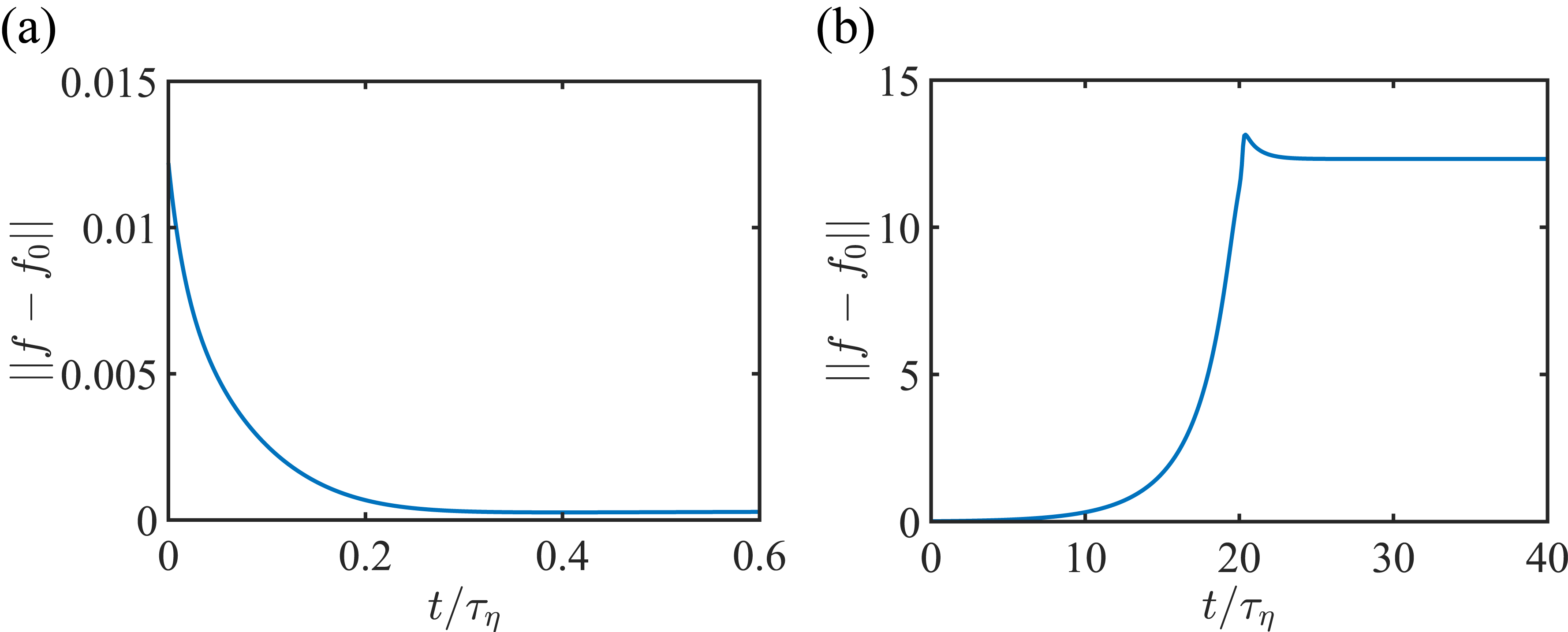}
\caption{\textbf{Dynamical classification of the stability of a stationary state.}
Evolution of the distance $\lVert f-f_0\rVert$ between the shape $f=(r,z)$ obtained from the time-dependent simulation and the stationary shape $f_0=(r_0,z_0)$ used as the initial condition, as defined in Eq.~(\ref{eq:disfun}). The simulation is initialized from the stationary state with the small random perturbation of Sec.~\ref{Appendix:perturbation}.
\textbf{(a)} Stable case: the distance decays back to zero, so the perturbed surface relaxes to the same stationary shape. Only the initial part of the time window is shown, so that the decay is resolved.
\textbf{(b)} Unstable case: the distance grows and the surface departs from the stationary state, settling onto a different branch. All stability simulations are run for a total duration of $40\,\tau_\eta$.}
\label{figS:stability}
\end{figure}

The stationary BVPs are solved with MATLAB's collocation solver \texttt{bvp5c}, using relative and absolute tolerances of $10^{-5}$ and $10^{-8}$, respectively, and a maximum mesh size of $5\times10^4$. Mirror-symmetric branches are initialized from equatorially symmetric profiles. Symmetry-broken branches are obtained by seeding the stationary solver with a small antisymmetric displacement. Local stability of a stationary solution is assessed dynamically. Denoting the stationary shape by $[r_0(u),z_0(u)]$, we take it as the initial condition, add the small random perturbation described in Sec.~\ref{Appendix:perturbation}, and evolve the time-dependent equations. Writing the evolving shape as $[r(u,t),z(u,t)]$, we monitor its distance from the stationary state,
\begin{equation}
\label{eq:disfun}
\mathrm{dis}(t)=\lVert r(u,t)-r_0(u)\rVert+\lVert z(u,t)-z_0(u)\rVert,
\end{equation}
where $\lVert f\rVert=\left(\int_0^1 f(u)^2\,du\right)^{1/2}$ is the $L^2$ norm of the function $f$. The stationary state is classified as stable when $\mathrm{dis}(t)$ decays with time, so that the perturbed surface relaxes back to the stationary shape, and unstable when $\mathrm{dis}(t)$ grows and the surface departs from it. Each classification run is integrated for a total duration of $40\,\tau_\eta$, which is long enough for the two behaviours to be distinguished unambiguously. Supplementary Fig.~\ref{figS:stability} shows $\mathrm{dis}(t)$ for a representative stable and a representative unstable case. The supercritical or subcritical character is determined from the direction and stability of the bifurcating asymmetric branch. The square-root scaling of the ring displacement near the bifurcation is shown in Supplementary Fig.~\ref{figS:supercritical}.

\subsection{Division observables and outcome classification}
The furrow position $u_{\mathrm{furrow}}$ is defined as the location where the meridional curvature $C^u_u$ is most negative. Its displacement from the equator is
\begin{equation}
\Delta u=u_{\mathrm{furrow}}-0.5.
\end{equation}
Once a narrow neck has formed, the volumes on the two sides of the furrow define the larger and smaller daughter volumes, $V_b$ and $V_s$, and the division-asymmetry index is
\begin{equation}
\chi=\frac{V_b-V_s}{V_b+V_s}.
\end{equation}
When a division proceeds, the furrow radius shrinks toward zero. The computation is stopped once the furrow radius $r_{\mathrm{furrow}}=\min_u r(u,t)$ falls below a numerical cutoff of $0.001\,R_0$, at which point the two sides of the neck are counted as separate daughter cells. Dynamic trajectories that constrict about the equator have $\chi\simeq0$ and are classified as symmetric division. Trajectories in which the furrow leaves the equator and produces unequal daughter volumes are classified as asymmetric division. Trajectories that instead relax to a weakly invaginated stationary state, without the furrow developing a narrow neck, are classified as failed division; these points are assigned $\chi=1$ only for visualization in the phase plots and do not represent a completed partition.

\subsection{Random perturbation of the initial state}\label{Appendix:perturbation}
To investigate how sensitively the division outcome depends on the initial condition, we apply a small random perturbation of the active stress at the first time step. Specifically, we replace the active couplings entering the constitutive law of Eq.~\eqref{eq:consti_Td} by
\begin{equation}
\label{eq:perturbation}
    \tilde{\xi}_s \to \tilde{\xi}_s + \sum_{n=1}^{10} a_n \cos(n\pi u), \qquad
    \tilde{\xi}_b \to \tilde{\xi}_b + \sum_{n=1}^{10} b_n \cos(n\pi u),
\end{equation}
where the coefficients are drawn independently for each mode as $a_n=r_n\tilde{\xi}_s$ and $b_n=r_n\tilde{\xi}_b$, with $r_n$ drawn from a uniform distribution on $[-0.01,\,0.01]$, so that both couplings receive the same relative perturbation. The modes $\cos(n\pi u)$ with odd $n$ break the mirror symmetry about the equator. From the second time step onward the perturbation is removed and the couplings revert to their unperturbed values $\tilde{\xi}_s$ and $\tilde{\xi}_b$, so that the subsequent evolution is governed by the unmodified equations and the perturbation acts purely on the initial condition. Each data point in Figs.~\ref{fig:size_confinement}A,E and~\ref{fig:regulation}B--E is obtained from ten independent realizations of this perturbation, and the error bars report the resulting standard deviation.

\suppnote{Experimental data analysis}

\subsection{Embryonic morphological datasets}
We analyze two published cell-resolved four-dimensional morphological datasets of \textit{C.~elegans} embryogenesis: the \textit{CMap} dataset, comprising eight mechanically uncompressed embryos, and the \textit{CShaper} dataset, comprising seventeen mechanically compressed embryos~\cite{Guan2025,cao2020}. Cell membranes are fluorescently labelled for three-dimensional segmentation and cell nuclei are labelled for lineage tracing. The stereotyped lineage nomenclature provides a common cell identity across embryos and permits the same division event to be compared between samples and mechanical conditions.

\subsection{Cell volumes and division asymmetry}
Volumes and division asymmetries are computed separately for each embryo. Within a given embryo, the segmented volume of each cell is averaged over all time points in that cell's lifespan. For each division event, the two averaged daughter volumes then give the experimental division-asymmetry index, using the same definition as in the simulations,
\begin{equation}
\chi=\frac{V_b-V_s}{V_b+V_s}.
\end{equation}
Every embryo therefore contributes one value of $\chi$ per division event, and the statistical unit in the analyses below is a single lineage-resolved division event in a single embryo rather than an individual image frame. The distributions shown in Fig.~\ref{fig:size_effect}B,D are taken over these per-embryo values.

In the box plots of Fig.~\ref{fig:size_effect}B,D, the box spans the first and third quartiles $Q_1$ and $Q_3$, the central line marks the median, and the whiskers extend to the most extreme values lying within $[Q_1-3\,\mathrm{IQR},\,Q_3+3\,\mathrm{IQR}]$, where $\mathrm{IQR}=Q_3-Q_1$ is the interquartile range. Values outside this range are plotted individually as outliers.

For the cell-size analysis in Fig.~\ref{fig:size_effect}B, somatic cell division events from the uncompressed dataset are grouped into four mother-cell-volume intervals: $10^1$--$10^2$, $10^2$--$10^3$, $10^3$--$10^4$, and $10^4$--$10^5\,\mu\mathrm{m}^3$. Adjacent volume groups are compared using a one-sided Wilcoxon rank-sum test. For the P-lineage analysis in Fig.~\ref{fig:size_effect}D, the division-asymmetry distributions of successive germline cell division events are compared between adjacent events using the same test. The annotations in Fig.~\ref{fig:size_effect}B,D use n.s. for $p>0.1$, one asterisk for $p<0.1$, two asterisks for $p<0.05$, and three asterisks for $p<0.001$.

\subsection{Compression comparison}
To quantify the confinement effect, division events are matched between the uncompressed and compressed datasets by their lineage identity. Each point in Fig.~\ref{fig:confinement_effect}B therefore represents one unique division with asymmetry $\chi_U$ in the uncompressed dataset and $\chi_C$ in the compressed dataset. Linear regression of the matched values gives the reported slope of $1.151$ and $R^2=0.917$; the identity line provides the reference for no compression-induced change.

\subsection{Cortex-contour comparison and lineage simulation}
For the P0 comparison, cell boundaries are obtained from automatic segmentation of the time-lapse images~\cite{segmentation}. Experimental and simulated contours are aligned in the same axial cross-section and compared at six matched stages spanning furrow initiation to near-complete division. Let $A$ be the set of pixels enclosed by the experimentally segmented contour and $B$ the set of pixels enclosed by the simulated contour. The similarity index is defined as the fraction of the experimental region that is recovered by the simulation,
\begin{equation}
\label{eq:similarity}
\mathrm{similarity}=\frac{|A\cap B|}{|A|},
\end{equation}
where $|\cdot|$ denotes the number of pixels in a set. The similarity index shown in Fig.~\ref{fig:validation}A remains above $90\%$ at all displayed stages.

The lineage calculation begins from a single calibrated P0 radius $\tilde R_0=5.4$ and uses the shared parameter set of Supplementary Table~\ref{tab:parameter} at fixed confinement $\epsilon=0.87$. Somatic blastomeres divide without the localized polarity profile, whereas the successive P-lineage blastomeres use the localized modulation of $\kappa(u)$. The two daughter volumes computed at each completed division become the mother-cell volumes for the next generation. Lineage termination is set by the cell-size-dependent phase boundary obtained from the bifurcation analysis. At the fixed activity and confinement used here, the critical size $\tilde R_{0,c}$ is defined by $\tilde{\xi}_s=\tilde{\xi}_s^*(\tilde R_{0,c},\epsilon)$, where the fixed-activity trajectory crosses the pitchfork boundary separating the asymmetric-division and failed-division regimes. For $\tilde R_0<\tilde R_{0,c}$ the mirror-symmetric branch is stable, so the cortex relaxes to a weakly invaginated stationary state without completing division and the corresponding branch is terminated. The boundary relevant to the somatic divisions is the one mapped in Fig.~\ref{fig:size_confinement}A,B, whereas the polarity-regulated germline divisions terminate at the failure boundary of Fig.~\ref{fig:regulation}D, which the localized $\kappa$-modulation shifts slightly.

\clearpage

\addcontentsline{toc}{section}{Supplementary Tables}

\begin{table}[H]
    \centering
    \begin{tabular}{cc}
    \hline
        Variation & \ Implied ODE\\         
        \hline
         $\partial_t r$ & $\alpha' = \cdots $ \\
         $\partial_t z$ & $\beta' = \cdots $ \\
         $\partial_t\psi$ & $\psi'' = \cdots$  \\
         $\partial_t h$ &    $\zeta' = \cdots $ \\
         $\partial_t\alpha$ & $r' = \cdots $ \\
         $\partial_t\beta$ & $z' = \cdots $   \\
         $\partial_t \zeta$ & $h' = 0\hspace{0.32cm}$      \\
         $v^u$ & $(v^u)'' = \cdots $ \\
         $v^{\phi}$ & $(v^{\phi})'' = \cdots $\\
    \hline
    \end{tabular}
    \caption{Structure of the ODE system with differential equations and their corresponding variations.}
    \label{table:ODE}
\end{table}

\begin{table}[H]
    \centering
    \begin{tabular}{ccc}
    \hline
        Variation  & \multicolumn{2}{c}{\ Boundary conditions}\\
        \hline
         $\partial_t r$ & $r(0) = 0,\,r(1) = 0$ \\
         $\partial_t z$ & $\beta(0) = 0,\,\beta(1) = 0$ \\
         $\partial_t\psi$ & $\psi(0) = 0$, $\psi(1) = \pi$ \\
         $\partial_t h$  & $\zeta(0) = 0,\,\zeta(1) = 0$ \\
         $v^u$ & $v^u(0) = 0,\,v^u(1) = 0$ \\
         $v^{\phi}$ & $v^{\phi}(0) = 0,\,v^{\phi}(1) = 0$ \\ 
    \hline
    \end{tabular}
    \caption{Boundary conditions arise from boundary terms of the variation Eqs.~(\ref{eq:Boundary_vu})-(\ref{eq:Boundary_psi}).}
    \label{Table:Boundry}
\end{table}

\begin{table}[H]
    \centering
    \begin{tabular}{cccc}
        \hline
        Parameter & Biological/Computational Meaning & Value & Unit \\
        \hline
        $\kappa$ & Bending rigidity & 1 & $\kappa_0$ \\
        $\eta_b$ & Bulk viscosity & 1 & $\eta_b$ \\
        $\eta_s$ & Shear viscosity & 1 & $\eta_b$ \\
        $L_h$ & Hydrodynamic length & 1 & $L_h$ \\
        $\tau_\eta$ & Time scale & $\eta_b L_h^2/\kappa_0$ & $\tau_\eta$ \\
        $R_0$ & Radius of the P0 zygote & 5.4 & $L_h$ \\
        $\xi_s\Delta\mu$ & Symmetric active coupling & 25 & $\kappa_0/L_h$ \\
        $\xi_b\Delta\mu$ & Bending active coupling & $-2.5$ & $\kappa_0/L_h$ \\
        $\Gamma$ & Friction coefficient & 1 & $\eta_b/L_h^2$ \\
        $\omega$ & Regulation profile width & 0.0577 & 1 \\
        $\Delta_\kappa$ & Regulation profile peak & 0.1 & 1 \\
        $u_0$ & Regulation position & 0.465 & 1 \\
        \hline
    \end{tabular}
    \caption{Parameters used in the model. This shared parameter set is used for the lineage-tree simulation of Fig.~\ref{fig:validation}B, in which $R_0$ is the radius of the P0 zygote and the cell size decreases from it at each subsequent division.}
    \label{tab:parameter}
\end{table}

\clearpage

\addcontentsline{toc}{section}{Supplementary Figures}

\begin{figure}[H]
\centering
\includegraphics[width = 0.8\columnwidth]{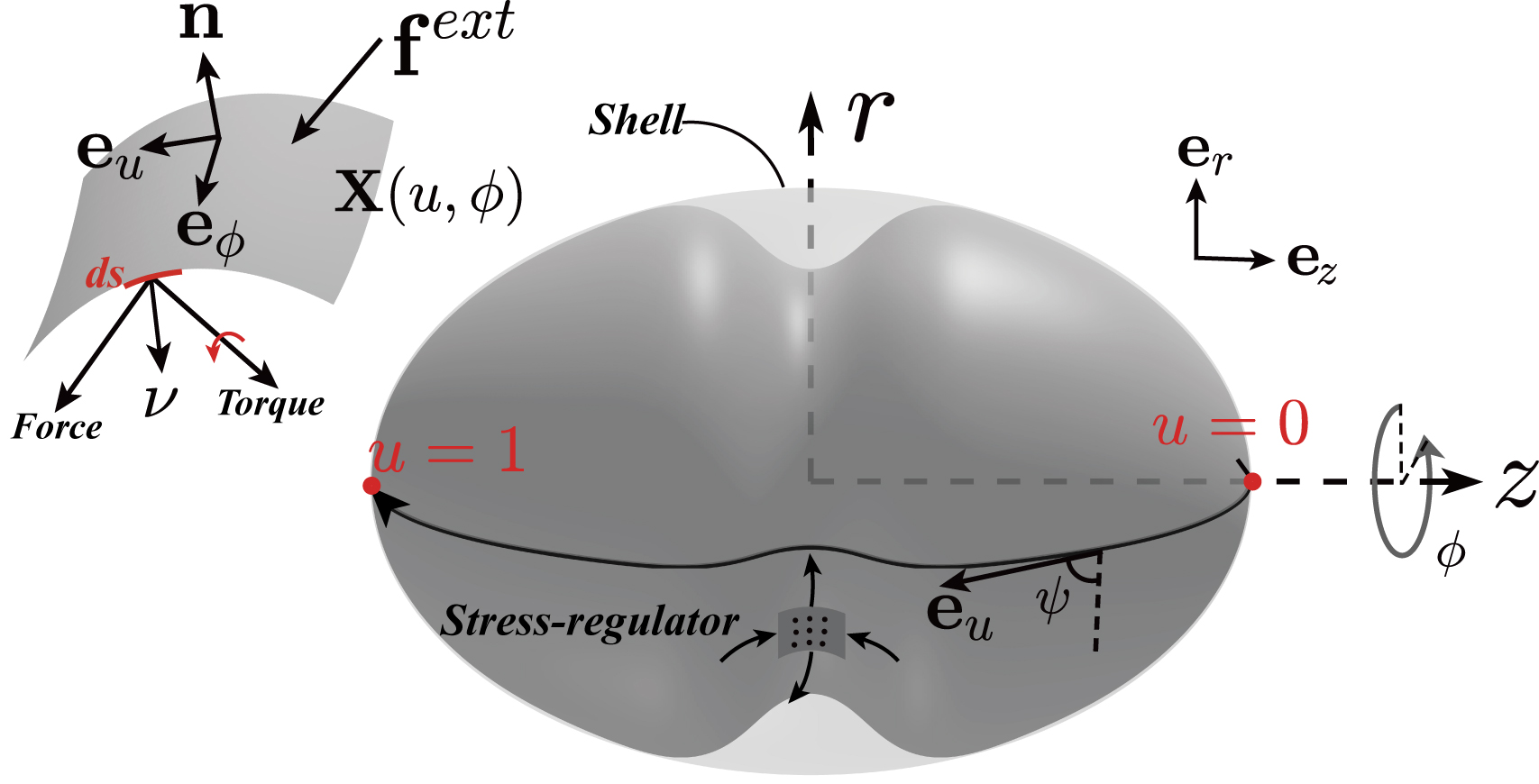}
\caption{Parameterization of axisymmetric surfaces $\mathbf{X(\mathit{u},\phi,t)}$. The force and torque are described by the tensors $t_{ij}$ and $m_{ij}$ acting on the line element $ds$. The geometric shell (semi-transparent gray ellipsoidal shell) is expressed as a potential well outside the enclosed active surface (dark gray opaque solid).}
\label{figS:Illustration}
\end{figure}

\begin{figure}[H]
\centering
\includegraphics[width=0.92\textwidth]{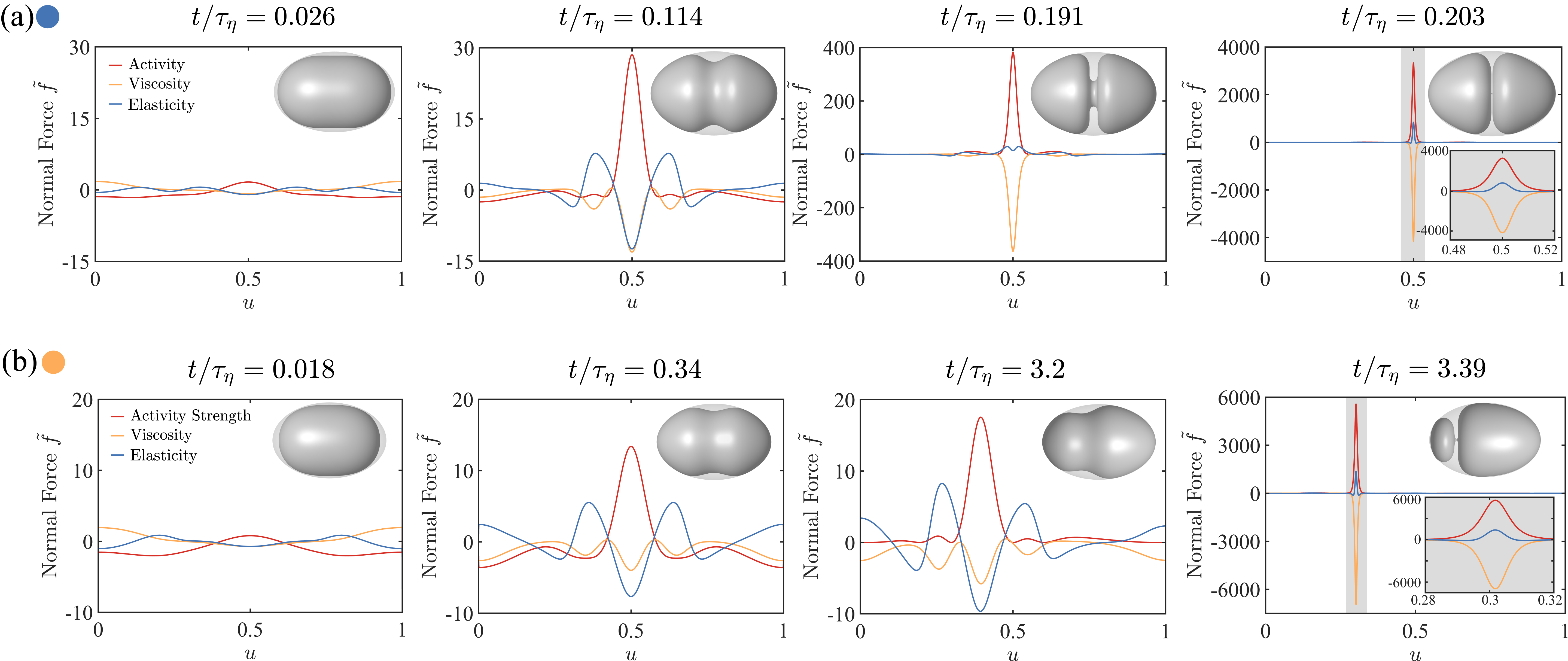}
\caption{\textbf{Normal force balance during symmetric and asymmetric division.}
\textbf{(a)} Symmetric division (blue, $\chi\approx 0$): the dimensionless normal force $\tilde{f}$ decomposed into active (red), viscous (orange), and elastic (blue) contributions, shown at four successive time points spanning the division process (insets: cortex shapes). The furrow forms symmetrically at $u=0.5$; the active force progressively concentrates at the equatorial region as constriction proceeds.
\textbf{(b)} Asymmetric division (orange): same force decomposition at four time points. The furrow displaces off the equatorial position; the active and viscous contributions develop an asymmetric spatial profile, and the final panel shows the near-pinched asymmetric furrow (inset, zoomed).}
\label{figS:forcebalance}
\end{figure}

\begin{figure}[H]
\centering
\includegraphics[width=0.92\textwidth]{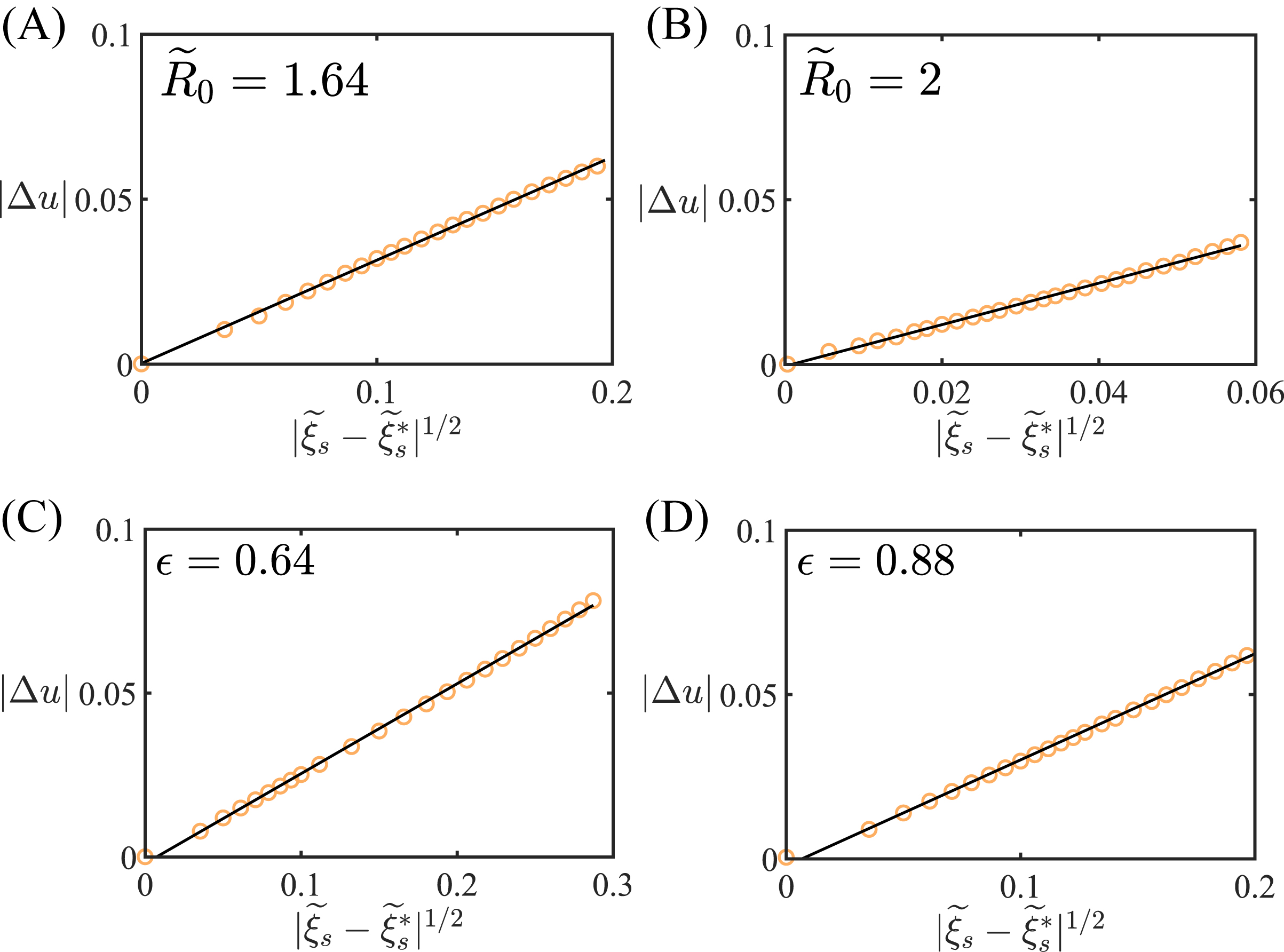}
\caption{\textbf{Square-root scaling of the symmetry-breaking bifurcation.}
Contractile-ring displacement $|\Delta u|$ plotted against $|\tilde{\xi}_s - \tilde{\xi}_s^*|^{1/2}$ for two cell radii (\textbf{A}, \textbf{B}: $\tilde{R}_0 = 1.64$, $2.00$; $\epsilon = 0.87$ fixed) and two confinement levels (\textbf{C}, \textbf{D}: $\epsilon = 0.64$, $0.88$; $\tilde{R}_0 = 2.22$ fixed). Orange circles: numerical data; black lines: linear fits. The linear relationship $|\Delta u| \propto |\tilde{\xi}_s - \tilde{\xi}_s^*|^{1/2}$ confirms that the transition is a pitchfork bifurcation in all cases shown; whether it is supercritical or subcritical is established separately from the branch stability in Fig.~\ref{fig:size_confinement}.}
\label{figS:supercritical}
\end{figure}

\begin{figure}[H]
\centering
\includegraphics[width=0.92\textwidth]{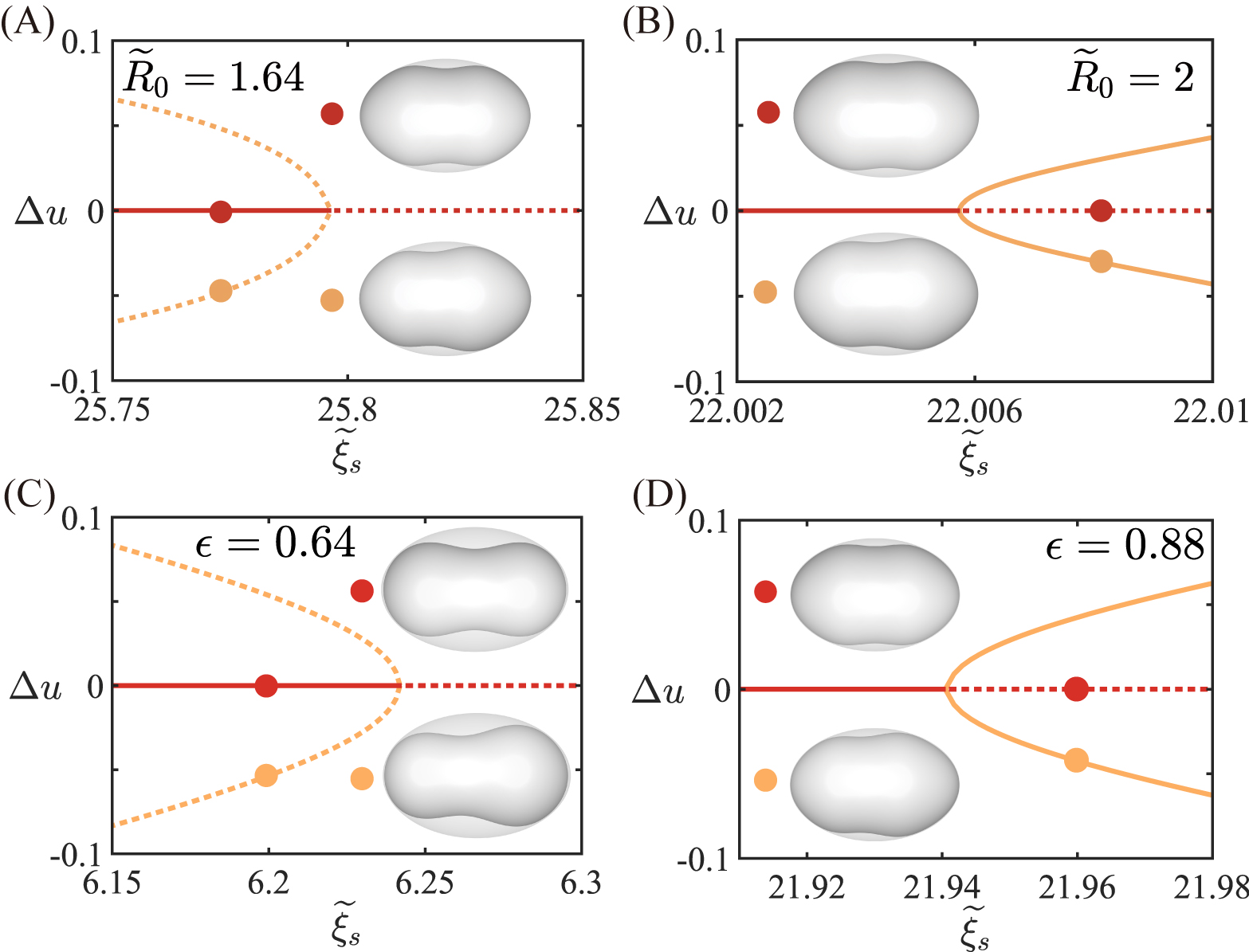}
\caption{\textbf{Near-threshold scaling of the contractile-ring displacement.}
Contractile-ring displacement $|\Delta u|=|u_\mathrm{furrow}-0.5|$ close to the symmetry-breaking threshold $\tilde{\xi}_s^*$, for the two control parameters examined in Fig.~\ref{fig:size_confinement}. Solid and dashed branches denote stable and unstable solutions, and the insets show representative cortex shapes.
\textbf{A,B}, Cell-size axis at fixed confinement $\epsilon=0.87$, for $\tilde{R}_0=1.64$ ($\tilde{\xi}_s^*=25.796$) and $\tilde{R}_0=2.00$ ($\tilde{\xi}_s^*=22.006$). The transition is subcritical at the smaller radius (\textbf{A}) and supercritical at the larger one (\textbf{B}).
\textbf{C,D}, Confinement axis at fixed size $\tilde{R}_0=2.22$, for $\epsilon=0.64$ ($\tilde{\xi}_s^*=6.24$) and $\epsilon=0.88$ ($\tilde{\xi}_s^*=21.94$). Here the ordering is reversed, the transition being subcritical at the weaker confinement (\textbf{C}) and supercritical at the tighter one (\textbf{D}).
In both sweeps $|\Delta u|$ follows the square-root scaling $|\Delta u|\propto|\tilde{\xi}_s-\tilde{\xi}_s^*|^{1/2}$ characteristic of a pitchfork bifurcation, but the character of the bifurcation shifts in opposite senses along the two control parameters.}
\label{figS:scaling}
\end{figure}

\vspace{0.5em}
\noindent\textbf{Supplementary Movie 1 $\vert$ Symmetric division at large cell size.}
Dynamic simulation of a large cell dividing symmetrically. Starting from the ellipsoidal rest shape imposed by the confining shell, activation of the curvature-dependent active stress nucleates a constriction furrow at the equatorial plane that deepens symmetrically, partitioning the cell into two daughters of equal volume. The dark gray surface is the active cortex and the semi-transparent gray ellipsoid is the confining shell; time is given in units of the viscous relaxation time $\tau_\eta$. This movie corresponds to the symmetric division of Fig.~\ref{fig:mechanical_model}C.\vspace{0.5em}
\noindent\textbf{Supplementary Movie 2 $\vert$ Asymmetric division via spontaneous symmetry breaking.}
Dynamic simulation of a cell in the symmetry-broken regime. The furrow initially forms near the equator but, as it deepens, spontaneously slips off the equatorial plane and migrates toward one pole, yielding two daughters of unequal volume. Rendering and time units as in Movie~S1. This movie corresponds to the asymmetric division of Fig.~\ref{fig:mechanical_model}D.\vspace{0.5em}
\noindent\textbf{Supplementary Movie 3 $\vert$ Failed division at reduced cell size.}
Dynamic simulation of a small cell. The cortex develops only a shallow, weakly invaginated furrow at the equatorial plane that fails to constrict to completion, so that cytokinesis does not proceed. Rendering and time units as in Movie~S1. This movie corresponds to the failed-division regime of Fig.~\ref{fig:size_confinement}.\vspace{0.5em}
\noindent\textbf{Supplementary Movie 4 $\vert$ Reproduction of the P0 zygote division.}
Dynamic simulation of the first (P0) cleavage of the \textit{C.\,elegans} embryo using the polarity-regulated model with a localized modulation of cortical bending rigidity. The furrow is biased slightly off the equator, reproducing the mildly asymmetric division that yields the larger AB and smaller P1 blastomeres. Rendering and time units as in Movie~S1. This movie corresponds to the P0 division of Fig.~\ref{fig:validation}A. \clearpage \renewcommand{\refname}{Supplementary References} \addcontentsline{toc}{section}{Supplementary References}

\clearpage

\end{document}